# Emotion self-regulation training in major depressive disorder using simultaneous real-time fMRI and EEG neurofeedback

Vadim Zotev[1#], Ahmad Mayeli[1,2], Masaya Misaki[1], Jerzy Bodurka[1,3#]

[1]Laureate Institute for Brain Research, Tulsa, OK, USA; [2]Electrical and Computer Engineering, University of Oklahoma, Tulsa, OK, USA; [3]Stephenson School of Biomedical Engineering, University of Oklahoma, Norman, OK, USA

**Abstract:** Simultaneous real-time fMRI and EEG neurofeedback (rtfMRI-EEG-nf) is an emerging neuromodulation approach, that enables simultaneous volitional regulation of both hemodynamic (BOLD fMRI) and electrophysiological (EEG) brain activities. Here we report the first application of rtfMRI-EEG-nf for emotion self-regulation training in patients with major depressive disorder (MDD). In this proof-of-concept study, MDD patients in the experimental group ($n$=16) used rtfMRI-EEG-nf during a happy emotion induction task to simultaneously upregulate two fMRI and two EEG activity measures relevant to MDD. The target measures included BOLD activities of the left amygdala (LA) and left rostral anterior cingulate cortex (rACC), and frontal EEG asymmetries in the alpha band (FAA, [7.5-12.5] Hz) and high-beta band (FBA, [21-30] Hz). MDD patients in the control group ($n$=8) were provided with sham feedback signals. An advanced procedure for improved real-time EEG-fMRI artifact correction was implemented. The experimental group participants demonstrated significant upregulation of the LA BOLD activity, FAA, and FBA during the rtfMRI-EEG-nf task, as well as significant enhancement in fMRI connectivity between the LA and left rACC. Average individual FAA changes during the rtfMRI-EEG-nf task positively correlated with depression and anhedonia severities, and negatively correlated with after-vs-before changes in depressed mood ratings. Temporal correlations between the FAA and FBA time courses and the LA BOLD activity were significantly enhanced during the rtfMRI-EEG-nf task. The experimental group participants reported significant mood improvements after the training. Our results suggest that the rtfMRI-EEG-nf may have potential for treatment of MDD.

**Keywords:** depression, neurofeedback, emotion regulation, amygdala, rostral anterior cingulate cortex, EEG-fMRI, frontal EEG asymmetry

## 1. Introduction

We have introduced simultaneous real-time fMRI and EEG neurofeedback (rtfMRI-EEG-nf) – a non-invasive neuromodulation approach, that enables simultaneous volitional regulation of both hemodynamic (BOLD fMRI) and electrophysiological (EEG) brain activities (Zotev et al., 2014). It involves real-time integration of concurrent fMRI and EEG data streams to provide real-time fMRI neurofeedback (rtfMRI-nf) and EEG neurofeedback (EEG-nf) signals simultaneously to a participant inside the MRI scanner (Mano et al., 2017; Zotev et al., 2014). This multimodal neurofeedback approach holds two major promises for treatment of neurological and psychiatric disorders. First, application of rtfMRI-EEG-nf may conceivably have stronger therapeutic effects than standalone applications of either rtfMRI-nf (e.g. Thibault et al., 2018) or EEG-nf (e.g. Micoulaud-Franchi et al., 2015). The reason is that rtfMRI-EEG-nf can target disorder-specific brain activity measures identified by two very different imaging modalities – fMRI and EEG (e.g. Mulert and Lemieux, 2010). In particular, relevant EEG measures can represent different EEG frequency bands, while BOLD fMRI activity reflects cumulative metabolic energy demands across the entire EEG spectrum. Second, rtfMRI-EEG-nf training may help to develop personalized mental strategies that would reliably engage both the fMRI and EEG target brain activities at the same time and further enhance their interactions. Such experimentally verified mental strategies could then be employed during EEG-nf-only training, which may provide a cost-effective, mobile, and long-term therapy in support of the rtfMRI-EEG-nf training. Until now, rtfMRI-EEG-nf has only been used in proof-of-principle studies with healthy participants (Perronnet et al., 2017; Zotev et al., 2014).

Here we report the first application of rtfMRI-EEG-nf for emotion self-regulation training in a neuropsychiatric population, specifically – in patients with major depressive disorder (MDD). During the rtfMRI-EEG-nf task, the participants induced happy emotion by retrieving and contemplating happy autobiographical memories, and, simultaneously, learned to upregulate two rtfMRI-nf signals and two EEG-nf signals. The four neurofeedback signals represented four brain activity measures relevant to MDD, as we explain below.

#Corresponding authors. E-mail: vzotev@laureateinstitute.org; jbodurka@laureateinstitute.org

The first rtfMRI-nf signal in our study is based on BOLD activity of the left amygdala (LA) target region of interest (ROI), as used in our previous rtfMRI-nf emotion self-regulation studies with healthy participants (Zotev et al., 2011), MDD patients (Young et al., 2014; Zotev et al., 2016), and PTSD patients (Zotev et al., 2018b). The amygdala plays a fundamental role in emotion processing. In MDD, the amygdala exhibits blunted BOLD responses to positive emotional stimuli (e.g. Price and Drevets, 2012; Suslow et al., 2010; Victor et al., 2010). Upregulation of the LA activity using the rtfMRI-nf during the positive emotion induction task based on retrieval of happy autobiographical memories has been shown to correct this amygdala reactivity bias in MDD, and lead to significant reduction in depression severity (Young et al., 2017). A similar rtfMRI-nf training procedure in PTSD patients yielded significant reductions in both PTSD severity and comorbid depression severity (Zotev et al., 2018b).

The second rtfMRI-nf signal is based on BOLD activity of a target ROI in the left rostral anterior cingulate cortex (rACC). The rACC function is very important in MDD (e.g. Pizzagalli, 2011). Resting rACC activity, measured by PET prior to antidepressant treatment, has been shown to reliably predict MDD patients' treatment response (Mayberg et al., 1997; Pizzagalli, 2011). After treatment, MDD patients exhibit enhanced fMRI functional connectivity between the rACC and the amygdala, both at rest and during exposure to neutral and positive emotional images (Anand et al., 2005). We have demonstrated that, in healthy participants, the rtfMRI-nf modulation of the LA activity during happy emotion induction is accompanied by significant enhancement, across nf runs, in fMRI connectivity between the LA and the left rACC (Zotev et al., 2011). Moreover, effective connectivity analyses suggest that the left rACC modulates activity of the LA and several prefrontal regions during the rtfMRI-nf training (Zotev et al., 2013). In MDD patients, resting fMRI connectivity between the LA and the left rACC shows negative correlation with depression severity (Yuan et al., 2014). This connectivity is increased after the rtfMRI-nf training of the LA activity (Yuan et al., 2014). Collectively, these results suggest that enhancement in fMRI connectivity between the LA and the left rACC by means of rtfMRI-nf during happy emotion induction should be beneficial to MDD patients.

Following this evidence, we included the rtfMRI-nf signal based on the left rACC activity to enable enhancement in fMRI connectivity between the LA and the left rACC through simultaneous upregulation of the two nf signals. We chose this approach, because it is the simplest way to enhance fMRI connectivity of the two brain regions, that does not involve computation of a correlation coefficient for their fMRI waveforms. A conceptually similar 'two-point' technique was independently used by Ramot et al., 2017. We did not expect a significant mean fMRI activation of the left rACC during the rtfMRI-EEG-nf task relative to rest, because the rACC is a part of the default mode network, and its mean activation during the rtfMRI-nf training of the LA activity is relatively low (Zotev et al., 2013).

The first EEG-nf signal in our study represents a change in frontal alpha EEG asymmetry, which we abbreviate here as FAA. The FAA is defined as $\ln(P(\text{right})) - \ln(P(\text{left}))$, where $P$ is EEG power in the alpha frequency band for corresponding (pre)frontal EEG channels on the right and on the left (e.g. F4 and F3). Because change in alpha EEG power negatively correlates with cortical neuronal activation (e.g. Cook et al., 1998), a more positive FAA indicates a relatively stronger activation of the left prefrontal regions. The FAA is commonly interpreted according to the approach-withdrawal hypothesis, which posits that activation of the left prefrontal regions is more closely associated with approach motivation, while activation of the right prefrontal regions is associated with avoidance motivation (e.g. Davidson, 1996; Harmon-Jones and Gable, 2018; Spielberg et al., 2011, 2013). Indeed, approach-related emotional states, such as happiness, are characterized by more positive FAA levels than avoidance-related states, such as fear or disgust (e.g. Davidson, 1996; Stewart et al., 2014). MDD patients and individuals with a history of depression show significantly lower FAA levels during an emotional task (either approach- or avoidance-related) than non-depressed participants performing the same task (Stewart et al., 2011, 2014). Resting-state FAA is also reduced in MDD patients compared to non-depressed individuals (e.g. Smith et al., 2018; Thibodeau et al., 2006), though the findings are less robust than those for an emotional challenge (Stewart et al., 2014). The task-related FAA results suggest that upregulation of FAA using EEG-nf during happy emotion induction would benefit MDD patients. Emotion regulation training with FAA-based EEG-nf has been explored in several studies (e.g. Allen et al., 2001; Baehr et al., 1997; Cavazza et al., 2014; Choi et al., 2011; Peeters et al., 2014; Quaedflieg et al., 2016; Rosenfeld et al., 1995).

Importantly, the FAA upregulation using the EEG-nf in the present experimental design is consistent with the LA BOLD activity upregulation using the rtfMRI-nf. Motivation is an important component of neurofeedback learning (e.g. Gaume et al., 2016). Because rtfMRI-nf training in general is a goal-oriented behavior, it requires approach motivation to be successful (e.g. Spielberg et al., 2011, 2013). In our previous study, MDD patients, who underwent rtfMRI-nf training of the LA activity, showed positive FAA changes, indicative of stronger approach motivation, during the rtfMRI-nf task (Zotev et al., 2016). Moreover, mean FAA changes correlated with the amyg-dala BOLD laterality values. Temporal correlation between the FAA time course and the LA BOLD activity was significantly enhanced during the rtfMRI-nf task (Zotev et al., 2016). These observations suggest that the FAA and the LA BOLD activity can be modulated simultaneously using the rtfMRI-EEG-nf.

The second EEG-nf signal represents a change in frontal high-beta EEG asymmetry, abbreviated here as FBA. The FBA is defined as $\ln(P(\text{left})) - \ln(P(\text{right}))$, where $P$ is EEG power in the high-beta (beta3) band [21-30] Hz for respective (pre)frontal EEG channels on the left and on the right (e.g. F3 and F4). Because cortical activation positively correlates with change in high-beta EEG power (Cook et al., 1998), a more positive FBA is associated with a relatively stronger activation of the left prefrontal regions (similar to FAA). MDD patients, when compared to healthy individuals, exhibit elevated resting high-beta EEG activity in the right prefrontal regions, and deficient high-beta activity in the precuneus/posterior cingulate (Pizzagalli et al., 2002). This finding suggests that resting-state FBA is reduced in MDD. Paquette et al. employed a high-beta EEG-nf for emotion self-regulation training in MDD patients (Paquette et al., 2009). The study showed that alleviation of MDD symptoms was associated with reduction in resting high-beta EEG activity in the prefrontal cortex and increase in such activity in the precuneus/posterior cingulate (Paquette et al., 2009). For responders in that study, the high-beta EEG activity reduction was larger in the right prefrontal regions than in the corresponding regions on the left (Paquette et al., 2009), indicating more positive resting-state FBA after the training. This finding suggests that more positive FBA may be beneficial to MDD patients. We have already demonstrated that healthy participants can learn to simultaneously upregulate the LA BOLD activity and the FBA using rtfMRI-EEG-nf while inducing happy emotion (Zotev et al., 2014).

We conducted the proof-of-concept rtfMRI-EEG-nf experiment, reported here, to test two main hypotheses. The first hypothesis was that MDD patients would be able to significantly increase the LA BOLD activity, the FAA, and the FBA using the rtfMRI-EEG-nf during happy emotion induction. As part of this hypothesis, we also predicted that fMRI connectivity between the LA and the left rACC would be significantly enhanced during the rtfMRI-EEG-nf task. The second hypothesis was that performance of the rtfMRI-EEG-nf task would be accompanied by significant enhancements in temporal correlations between the FAA and FBA time courses and the LA BOLD activity. Such enhancements would indicate that the corresponding EEG-nf and rtfMRI-nf signals were indeed upregulated together in real time, rather than independently one at a time (Zotev et al., 2014). The primary outcome measures in our study are the LA activation and the FAA upregulation. The secondary outcome measures are the left rACC vs LA functional connectivity enhancement and the FBA upregulation. In addition to testing the two hypotheses, we conducted exploratory analyses to examine associations between the target activity measures and the MDD patients' depression severity, anhedonia severity, and mood rating changes.

## 2. Methods

### 2.1. Participants

The study was conducted at the Laureate Institute for Brain Research. It was approved by the Western Institutional Review Board (IRB). All study procedures were performed in accordance with the principles expressed in the Declaration of Helsinki.

Participants were recruited through online, newspaper, radio, flyer, and other media advertisements. Individuals were recruited, if they were currently depressed, but not currently taking psychiatric medication. All participants underwent screening evaluations, including the Structured Clinical Interview for the Diagnostic and Statistical Manual of Mental Disorders, Fourth Edition (DSM-IV) (American Psychiatric Association, 2000) Axis I disorders. The following exclusion criteria were applied: general MRI exclusions, current pregnancy, psychosis, serious suicidal ideation, major medical or neurological disorders, exposure to any medication likely to influence cerebral function or blood flow within 3 weeks (8 weeks for fluoxetine), and meeting the DSM-IV criteria for drug or alcohol abuse within the previous year or for lifetime alcohol or drug dependence (except nicotine). Participants enrolled in the study met the criteria for MDD laid out in the DSM-IV. Most participants had recurrent MDD. They provided a written informed consent as approved by the IRB, and received monetary compensation.

Twenty four unmedicated MDD patients completed an rtfMRI-EEG-nf training session. Prior to the session, the participants underwent a psychological evaluation by a licensed psychiatrist. It included administration of the following tests: the 21-item Hamilton Depression Rating Scale (HDRS, Hamilton, 1960), the Montgomery-Asberg Depression Rating Scale (MADRS, Montgomery and Asberg, 1979), the Snaith-Hamilton Pleasure Scale (SHAPS, Snaith et al., 1995), the Hamilton Anxiety Rating Scale (HARS, Hamilton, 1959), the 20-item Toronto Alexithymia Scale (TAS-20, Bagby et al., 1994), and the Behavioral Inhibition System / Behavioral Activation System scales (BIS/BAS, Carver and White, 1994). Both before and after the rtfMRI-EEG-nf session, the participants completed the Profile of Mood States (POMS, McNair et al., 1971), the State-Trait Anxiety Inventory (STAI, Spielberger et al., 1970), and the Visual Analogue Scale (VAS) with 10-point subscales for happy, restless, sad, anxious, irritated, drowsy, and alert states. Three MDD patients were in remission on the day of the experiment (HDRS ratings ≤ 7). BIS/BAS scores were unavailable for three participants out of 24.

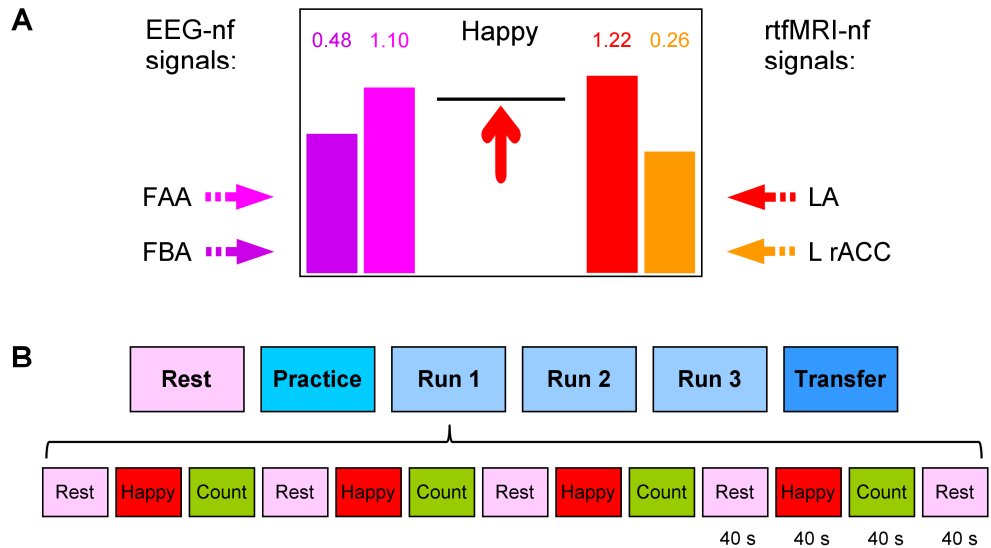

**Figure 1.** Experimental paradigm for emotion self-regulation training using simultaneous real-time fMRI and EEG neurofeedback (rtfMRI-EEG-nf). A) Real-time GUI display screen for Happy Memories conditions with rtfMRI-EEG-nf. The four neurofeedback signals are displayed on the screen as four variable-height bars. The two EEG-nf signals on the left are based, respectively, on changes in frontal alpha EEG asymmetry (FAA, magenta) and frontal high-beta EEG asymmetry (FBA, purple). The two rtfMRI-nf signals on the right are based, respectively, on fMRI activities of the left amygdala (LA, red) and the left rostral anterior cingulate cortex (L rACC, orange). The bar heights are updated every 2 s. The black horizontal bar in the middle of the screen specifies a target level. It is raised from run to run. B) Experimental protocol consisted of six runs, each lasting 8 min 46 s. It included a Rest run, four rtfMRI-EEG-nf runs – Practice, Run 1, Run 2, Run 3, and a Transfer run without nf. The names of the five task runs are abbreviated in the text and figures as PR, R1, R2, R3, and TR, respectively. The task runs consisted of 40-s long blocks of Rest, Happy Memories, and Count conditions. The condition names are abbreviated as R, H, and C, respectively. No bars were displayed during the Rest and Count conditions, and during the entire Transfer run.

The participants were assigned to either an experimental group (EG) or a control group (CG) at 2:1 ratio in numbers, common in proof-of-concept rtfMRI-nf studies (Young et al., 2014). All the participants were given identical instructions and were unaware of their group status. During the training session, participants in the EG ($n$=16, 13 females) received the rtfMRI-EEG-nf, based on their real-time EEG and fMRI brain activity measures. Participants in the CG ($n$=8, 4 females) were provided, without their knowledge, with sham feedback signals, unrelated to brain activity. Because the recruitment of unmedicated MDD patients was slow, the experiments for the EG were performed first, followed by the experiments for the CG. Psychological trait measures for the EG and CG participants, assessed before the rtfMRI-EEG-nf session, are reported in *Supplementary material* (Table S1). There were no significant group differences in these measures.

### 2.2. Real-time fMRI and EEG neurofeedback

The rtfMRI-EEG-nf was implemented using the custom real-time control system for integration of simultaneously acquired EEG and fMRI data streams, described in Zotev et al., 2014. The neurofeedback information was displayed to a participant inside the scanner on a projection screen via a multimodal graphical user interface (mGUI), depicted in Fig. 1A. The mGUI included four thermometer-style variable-height bars. The heights of these bars, updated every 2 s, represented the four neurofeedback signals. Each bar height was also indicated by a numeric value shown above that bar (Fig. 1A).

The red rtfMRI-nf bar on the right represented BOLD fMRI activity of the left amygdala (LA) target ROI (Fig. 2A). This spherical ROI with $R$=7 mm was centered at (−21, −5, −16) locus in the Talairach space (Talairach and Tournoux, 1988), as in our previous studies (Zotev et al., 2011, 2014, 2016). The orange rtfMRI-nf bar represented fMRI activity of the left rACC target ROI (Fig. 2B). This ROI, also with $R$=7 mm, was centered at (−3, 34, 5) locus, which had exhibited significant enhancements in both functional and effective fMRI connectivities with the LA during the rtfMRI-nf training in healthy participants (Zotev et al., 2011, 2013).

The magenta EEG-nf bar on the left represented a change in relative alpha EEG asymmetry for channels F3 and F4 (Fig. 2C). The relative alpha asymmetry ($A$) was defined as $A = (P(F4) - P(F3)) / (P(F4) + P(F3))$, where $P$ is EEG power in the alpha frequency band [7.5-12.5] Hz. Note that normalized frontal alpha EEG asymmetry, commonly defined as FAA = $\ln(P(F4)) - \ln(P(F3))$, has a Gaussian distribution, appropriate for statistical analyses. However, its infinite variation range makes it less convenient for real-time applications. Therefore, we employed a change in the relative asymmetry $A$ as a target measure for EEG-nf, and used the FAA in offline data analyses. The relative asymmetry $A$ and the normalized asymmetry FAA are related by the Fisher transform with factor ½ (Zotev et al., 2014). Similarly, the purple EEG-nf bar represented a change in relative high-beta EEG asymmetry ($B$), defined as $B = (P(F3) - P(F4)) / (P(F3) + P(F4))$, where $P$ is EEG power in the high-beta frequency band [21-30] Hz. Normalized FBA = $\ln(P(F3)) - \ln(P(F4))$ was used in offline data analyses. Thus, the rtfMRI-EEG-nf was designed to enable upregulation of both the FAA and FBA simultaneously with upregulation of BOLD activities of the LA and L rACC.

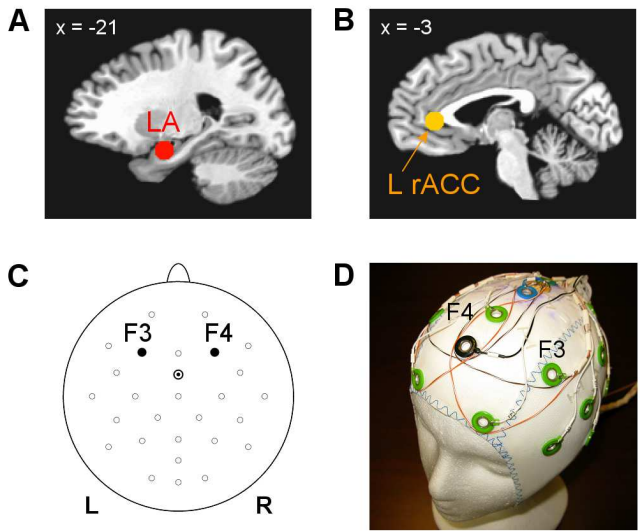

**Figure 2.** Target regions of interest (ROIs) and EEG channels used to provide the rtfMRI-EEG-nf. The ROIs were defined anatomically in the co-planar stereotaxic atlas of the human brain by Talairach and Tournoux, and transformed to each participant's individual fMRI image space. A) Spherical 14-mm-diameter target ROI in the left amygdala (LA) region, projected onto the standard TT_N27 anatomical template in the Talairach space. B) Spherical 14-mm-diatemer target ROI in the left rostral anterior cingulate cortex (L rACC) region. C) Frontal EEG channels F3 (left) and F4 (right) used to provide the EEG-nf based on frontal EEG asymmetries in the alpha and high-beta EEG bands. Channel FCz was employed as reference. D) Custom modification of a standard MR-compatible 32-channel EEG cap to improve quality of the EEG-nf during fMRI. The modified cap includes four wire contours with resistors for acquisition of reference artifact waveforms.

For the control group (CG), the four actual neurofeedback signals were substituted, without the participants' knowledge, with sham feedback signals, which were computer generated and unrelated to any brain activity. The sham feedback signals were computed, for each 40-s-long condition block, as random linear combinations of seven Legendre polynomials, as described previously (Zotev et al., 2018a). They were used to set heights of the four nf bars in real time (Fig. 1A). These signals' waveforms were smooth, but randomly shaped, and they also varied randomly across condition blocks and across participants.

### 2.3. Experimental protocol

The experimental protocol for training of emotion self-regulation using the rtfMRI-EEG-nf is illustrated in Fig. 1B. It has the same overall structure as the protocols we used previously (Zotev et al., 2011, 2014, 2016). The protocol included six EEG-fMRI runs (Fig. 1B), each lasting 8 min 46 s. During the Rest run, the participants were asked to relax and rest while looking at a fixation cross. The five task runs – the Practice run, Run 1, Run 2, Run 3, and the Transfer run – consisted of alternating 40-s-long blocks of Happy Memories, Count, and Rest conditions (Fig. 1B). Each condition was specified by visual cues that included a color symbol at the center of the screen and a text line at the top of the screen. For the Happy Memories with rtfMRI-EEG-nf condition blocks, the participants were asked to induce happy emotion by recalling happy autobiographical memories, while simultaneously trying to raise the levels of all four neurofeedback bars on the screen (Fig. 1A). For the Count condition blocks, the participants were instructed to mentally count back from 300 by subtracting a given integer. For the Rest condition blocks, the participants were asked to relax and rest while looking at a fixation cross. No bars were displayed during the Rest and Count conditions, and during the Happy Memories conditions in the Transfer run.

Prior to the experiment, the participants were asked to think of and write down three happy autobiographical memories, keeping them confidential. The Practice run was included to give the participants an opportunity to become familiar with the rtfMRI-EEG-nf procedure and evaluate emotional impact of the prepared happy memories. During the four Happy Memories condition blocks in the Practice run, the instruction line at the top of the screen read "Happy – try 1st memory", "Happy – try 2nd memory", "Happy – try 3rd memory", and "Happy – try best memory", respectively. During the subsequent training runs (Run 1, Run 2, Run 3), the participants were encouraged to use any (and as many) memories that helped them induce happy emotion and raise the rtfMRI-EEG-nf bars. The instruction cue during the Happy Memories condition blocks in these runs was "Happy" (Fig. 1A). The Transfer run without nf was included to evaluate whether the participants' learned ability to control the four target measures of brain activity generalized beyond the rtfMRI-EEG-nf training. During the Happy Memories condition blocks in the Transfer run, the instruction line read "As Happy as possible".

A target level for the rtfMRI-nf and EEG-nf signals was specified by the black horizontal bar above the red arrow in the middle of the mGUI screen (Fig. 1A). To encourage the participants to improve their performance from run to run, the target level was raised in a linear fashion across the four nf runs. For the rtfMRI-nf signals, the target bar heights corresponded to 0.5%, 1.0%, 1.5%, and 2.0% fMRI percent signal changes for the Practice run, Run 1, Run 2, and Run 3, respectively. For the EEG-nf signals, the same bar heights corresponded to $A$ and $B$ relative asymmetry changes by 0.05, 0.10, 0.15, and 0.20, respectively. The display ranges were from −3% to +3% for the fMRI-nf signals, and from −0.3 to +0.3 for the EEG-nf signals. The Count conditions involved counting back from 300 by subtracting 3, 4, 6, 7, and 9 for the Practice run, Run 1, Run 2, Run 3, and the Transfer run, respectively. After each experimental run with the Happy Memories condition, a participant was asked to rate his/her performance on a scale from 0 ("not at all") to 10 ("extremely") by verbally answering two questions: "How successful were you at recalling your happy memories?" and "How happy are you right now?".

## 2.4. MRI and EEG data acquisition

All experiments were conducted on the General Electric Discovery MR750 3T MRI scanner with a standard 8-channel receive-only head coil. A single-shot gradient echo EPI sequence with FOV/slice=240/2.9 mm, $TR/TE$=2000/30 ms, flip angle=90°, 34 axial slices per volume, slice gap=0.5 mm, SENSE $R$=2 in the phase encoding (anterior-posterior) direction, acquisition matrix 96×96, sampling bandwidth=250 kHz, was employed for fMRI. Each fMRI run included 263 EPI volumes (the first three EPI volumes were excluded from data analyses). Physiological pulse oximetry and respiration waveforms were recorded simultaneously with fMRI. The EPI images were reconstructed into a 128×128 matrix, resulting in 1.875×1.875×2.9 mm$^3$ fMRI voxels. A T1-weighted 3D MPRAGE sequence with FOV/slice=240/1.2 mm, $TR/TE$=5.0/1.9 ms, $TD/TI$=1400/725 ms, flip angle=10°, 128 axial slices per slab, SENSE $R$=2, acquisition matrix 256×256, sampling bandwidth=31.2 kHz, scan time=4 min 58 s, was used for structural imaging. It provided high-resolution anatomical brain images with 0.94×0.94×1.2 mm$^3$ voxels.

EEG recordings were performed simultaneously with fMRI using a 32-channel MR-compatible EEG system from Brain Products, GmbH. The system included one BrainAmp MR plus amplifier. The EEG system's clock was synchronized with the MRI scanner's 10 MHz clock using the Brain Products' SyncBox device. EEG data were acquired with 0.2 ms temporal and 0.1 μV measurement resolution (16-bit 5 kS/s sampling) in [0.016-250] Hz frequency band with respect to FCz reference. The ECG waveform was acquired with 0.5 μV resolution. BrainVision Recorder software was used for acquisition of raw EEG data, while BrainVision RecView software (Brain Products, GmbH) was employed for real-time EEG-fMRI artifact correction as described below. Technical details of the EEG-fMRI system setup, configuration, and raw data acquisition were described previously (Zotev et al., 2012).

## 2.5. Real-time EEG-fMRI artifact correction

In the present study, we implemented a novel procedure for more efficient real-time EEG-fMRI artifact correction to improve quality of EEG-nf during fMRI. It involved a special modification of a standard 32-channel MR-compatible EEG cap (BrainCap-MR from EASYCAP, GmbH). The cap modification is shown in Fig. 2D and described in detail in *Supplementary material* (S1.1). It enabled acquisition of four reference artifact waveforms, which we refer to as $R_1(t)$, $R_2(t)$, $R_3(t)$, and $R_4(t)$, approximating cardioballistic (CB) and random-motion artifacts picked up by EEG channels F3 and F4.

The real-time procedure for EEG-fMRI artifact correction is depicted schematically in Fig. 3A. It is implemented in BrainVision RecView software, which receives raw EEG data from BrainVision Recorder software in real time. The procedure includes three consecutive steps.

First, the RecView MRI Artifact Filter is used to perform real-time average artifact subtraction (AAS) of MR artifacts. The AAS method takes advantage of temporal periodicity of an fMRI pulse sequence (period=$TR$) and associated MR artifacts (Allen et al., 2000). After the correction, the data are lowpass filtered at 80 Hz (96 dB/octave) and downsampled to 250 S/s sampling rate (4 ms interval).

Second, real-time linear regression of CB and random-motion artifacts (Masterton et al., 2007) is conducted using the RecView Linear Derivation Filter. It is performed for channels F3 and F4 according to the formulas: $V_C(t,F3) = V(t,F3) − a_1R_1(t) − a_2R_2(t) − a_3R_3(t) − a_4R_4(t)$, and $V_C(t,F4) = V(t,F4) − b_1R_1(t) − b_2R_2(t) − b_3R_3(t) − b_4R_4(t)$. Here, $V_C(t,F3)$ and $V_C(t,F4)$ are corrected waveforms for F3 and F4 after the regression, $\{R_i(t)\}$, $i$=1…4, are the reference artifact waveforms, and $\{a_i\}$, $\{b_i\}$, $i$=1…4, are linear regression coefficients. The coefficients are determined before each experimental run as explained below.

Third, the RecView Pulse Artifact Filter is used to carry out AAS of CB artifacts. The AAS in this case relies on quasi-periodic nature of cardiac activity and related CB artifacts (Allen et al., 1998). Cardiac epochs are determined from the ECG waveform, and a moving average over 21 epochs is subtracted from each channel's data. Note that the linear regression procedure attenuates CB and random-motion artifacts without any assumptions about their temporal periodicity. Therefore, the linear regression and the AAS reduce CB artifacts independently, enabling more efficient real-time CB artifact suppression for channels F3 and F4.

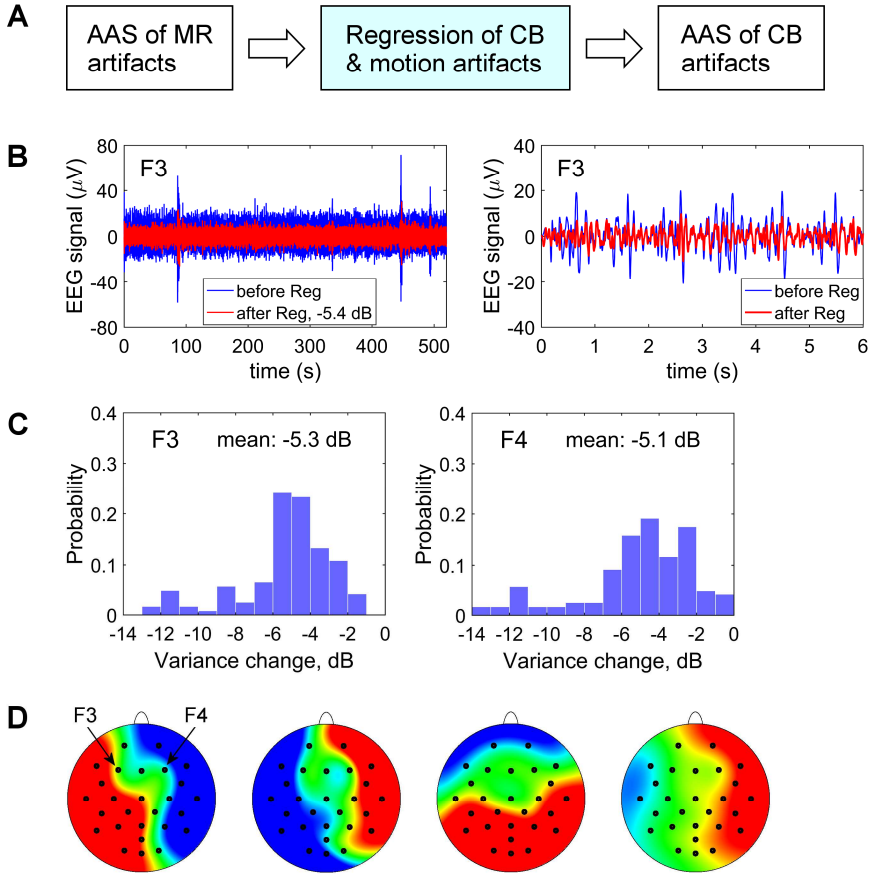

**Figure 3.** Real-time EEG-fMRI artifact correction procedure and its performance. A) The real-time procedure included three consecutive steps implemented in BrainVision RecView software: average artifact subtraction (AAS) of MR artifacts; regression of reference artifact waveforms approximating cardioballistic (CB) and random-motion artifacts (channels F3 and F4 only); AAS of cardioballistic artifacts. B) Illustration of performance of the regression procedure across an entire experimental run (left) and for a short time interval in the middle of that run (right). C) Performance of the regression procedure for channels F3 and F4 across all experimental (task) runs for all participants. Each histogram bar represents a probability of observing a change in signal variance within a given 1 dB-wide interval after the regression across one experimental run. D) Illustration of topographic properties of residual CB and random-motion artifacts after the real-time procedure (A). Topographies of independent components modeling such artifacts for a typical experimental run are shown. Note additional suppression of the artifacts for channels F3 and F4 compared to the surrounding EEG channels.

The linear regression coefficients $\{a_i\}$, $\{b_i\}$, $i=1…4$, are determined as follows. After the real-time application of the MRI Artifact Filter, the data are bandpass filtered in [5-35] Hz frequency range (96 dB/octave) using the RecView Frequency Filter, and saved to a file. After each experimental run, a MATLAB script is executed offline. It includes the glmfit() function to solve general linear models (GLMs) fitting the reference artifact waveforms to the waveforms from channels F3 and F4: $V(t,F3) = a_1R_1(t) + a_2R_2(t) + a_3R_3(t) + a_4R_4(t) + e(t,F3)$, and $V(t,F4) = b_1R_1(t) + b_2R_2(t) + b_3R_3(t) + b_4R_4(t) + e(t,F4)$. Here, $e(t,F3)$ and $e(t,F4)$ are neuronal and other signal components showing no correlations with the reference artifact waveforms. The fitting is carried out across the entire run. The GLM coefficients $\{a_i\}$, $\{b_i\}$, $i=1…4$, are then entered into the Linear Derivation Filter to enable the real-time artifact regression during the next experimental run. Thus, the real-time regression during the Practice run employs the coefficients determined from the data for the Rest run, the regression during Run 1 utilizes the coefficients computed from the data for the Practice run, and so on (Fig. 1B). Remarkably, the fact that the regression coefficients were determined from the preceding run's data had little effect on the efficiency of the artifact regression in terms of signal variance reduction, as reported in *Supplementary material* (S1.2, Fig. S1). This finding suggests that overall properties of CB and random-motion artifacts did not generally change much from one run to the next.

Performance of the real-time artifact regression procedure is demonstrated in Figs. 3B,C,D. Fig. 3B illustrates reduction in variance of an artifact-contaminated EEG signal after the regression for a typical experimental run. Histograms of signal variance changes across all task runs for all participants in both groups (24×5 = 120 runs) are shown in Fig. 3C. The mean signal variance reductions after the real-time regression across one run are 5.3 dB for F3 and 5.1 dB for F4. Larger variance reductions are observed for participants with stronger heart beats (taller and sharper R peaks in the ECG) leading to stronger CB artifacts in the alpha and beta EEG bands. Fig. 3D shows topographies of residual CB and random-motion artifacts from an independent component analysis (ICA) applied offline to representative single-run EEG data after the real-time artifact correction procedure (Fig. 3A). The inclusion of the real-time artifact regression for channels F3 and F4 led to an additional suppression of residual CB and random-motion artifacts for these two channels compared to surrounding EEG channels (Fig. 3D), for which only AAS of MR and CB artifacts were performed. Therefore, the described approach can present a practical, intuitive, and easy-to-use alternative to more advanced methods, such as real-time ICA (e.g. Mayeli et al., 2016).

*2.6. Real-time data processing*

Implementation of the rtfMRI-EEG-nf in the present study was similar to that in our previous work (Zotev et al., 2014), except that two rtfMRI-nf signals and two EEG-nf signals were computed and displayed to a participant at the same time (Fig. 1A).

Each rtfMRI-EEG-nf experiment began with acquisition of a high-resolution MPRAGE anatomical brain image, followed by acquisition of a short EPI dataset (5 volumes). The last volume of the EPI dataset was employed as a reference EPI volume defining the subject's individual EPI space. The MPRAGE image was transformed to the Talairach space, and this transformation was used as a template to transform the LA and L rACC target ROIs from the Talairach space (Figs. 2A,B) to the individual EPI space. The resulting ROIs in the EPI space contained approximately 140 voxels each. During the subsequent EEG-fMRI runs, the real-time plugin in AFNI (Cox, 1996; Cox and Hyde, 1997) was used to perform volume registration (Cox and Jesmanowicz, 1999) of each acquired EPI volume to the reference EPI volume and export mean values of fMRI signals for these two ROIs in real time. These fMRI signal values were sent to the mGUI software via a TCP/IP socket (Fig. 1 in Zotev et al., 2014). An rtfMRI signal for the LA target ROI was computed for each Happy Memories condition as a percent signal change with respect to the baseline obtained by averaging the LA fMRI signal values for the preceding 40-s-long Rest condition block (Fig. 1B). A moving average of the current and two preceding LA rtfMRI signal values was computed to reduce effects of fMRI noise and physiological artifacts. This moving average was used to set the height of the red rtfMRI-nf bar (Fig. 1A) every $TR$=2 s. An rtfMRI signal for the L rACC target ROI was calculated in the same way, and its moving average was used to set the height of the orange rtfMRI-nf bar (Fig. 1A).

During each EEG-fMRI run, the real-time correction of MR, random-motion, and CB artifacts was performed in the RecView software as described above (Fig. 3A). The corrected EEG data were exported in real time as data blocks of 8 ms duration via a TCP/IP socket to the EEG processing modules of the EEG-fMRI data integration software (Fig. 1 in Zotev et al., 2014). These modules were written in Python and utilized NumPy functions. FFT power spectra for channels F3 and F4 were computed every 2 s for a moving data interval of 2.048 s duration with Hann window. The relative alpha EEG asymmetry $A$ and the relative high-beta EEG asymmetry $B$ were calculated as described above. The $A$ and $B$ values were sent via a TCP/IP socket to the mGUI software, where they were processed along with the corresponding fMRI signal values using a separate software thread. For each Happy Memories condition, a change in $A$ was determined as a difference between the current $A$ value and the baseline obtained by averaging $A$ values for the preceding Rest condition block (Fig. 1B). A moving average of the current and two preceding $A$ changes was computed. This moving average (multiplied by 10) was used to set the height of the magenta EEG-nf bar (Fig. 1A) every 2 s. A change in $B$ for each Happy Memories condition was calculated in the same way, and its moving average (multiplied by 10) was used to set the height of the purple EEG-nf bar on the screen (Fig. 1A). The real-time $A$ and $B$ values and changes are compared to the corresponding $A$ and $B$ values and changes, determined in the offline EEG data analysis, in *Supplementary material* (S1.3, Fig. S2).

*2.7. fMRI data analysis*

Offline analysis of the fMRI data was performed in AFNI as described in detail in *Supplementary material* (S1.4). The analysis involved fMRI pre-processing with despiking, cardiorespiratory artifact correction (Glover et al., 2000), slice timing correction, and volume registration. A general linear model (GLM) fMRI activation analysis with Happy Memories and Count block-stimulus conditions was applied to the preprocessed fMRI data. Average GLM-based fMRI percent signal changes for the Happy Memories vs Rest condition contrast and for the Happy Memories vs Count contrast were computed for the LA and L rACC target ROIs (Figs. 2A,B) and used to characterize the rtfMRI-nf performance. For exploratory analyses, the amygdala BOLD laterality and the middle frontal gyrus (MidFG) BOLD laterality were computed as described in *Supplementary material* (S1.4).

Statistical results in the present study were corrected for multiple comparisons by controlling the False Discovery Rate (FDR $q$). In whole-brain fMRI and EEG-fMRI analyses, FDR correction was applied voxel-wise.

*2.8. fMRI-based PPI analysis*

To evaluate changes in the left amygdala fMRI functional connectivity between experimental conditions, we conducted a psychophysiological interaction (PPI) analysis (Friston et al., 1997; Gitelman et al., 2003), as described in detail in *Supplementary material* (S1.5). The analysis was based on fMRI time course for the LA seed ROI. This time course was used to define two PPI regressors: the fMRI-based PPI correlation regressor and the fMRI-based PPI interaction regressor for the Happy

Memories vs Rest condition contrast (S1.5). We selected this contrast a priori, because our earlier analyses (unpublished) had shown that the PPI interaction effects for the LA time course were more significant for the Happy Memories vs Rest contrast than for the Happy Memories vs Count contrast. A single-subject fMRI-based PPI analysis for each run involved fitting a GLM model with these two PPI regressors (in addition to other fMRI regressors, see S1.5).

*2.9. EEG data analysis*

Offline analysis of the EEG data was performed using BrainVision Analyzer 2.1 software (Brain Products, GmbH) as described in detail in *Supplementary material* (S1.6). Removal of EEG artifacts was based on the AAS (Allen et al., 1998, 2000) and independent component analysis (Bell and Sejnowski, 1995), implemented in Analyzer 2.1. Time-frequency analysis with Morlet wavelets was used to compute EEG power as a function of time and frequency. The upper alpha EEG frequency band was defined individually for each participant as [IAF, IAF+2] Hz, where IAF is the individual alpha peak frequency. The IAF was determined by inspection of average EEG spectra for the occipital and parietal EEG channels across the Rest condition blocks in the four nf runs (Fig. 1B). The normalized FAA was computed as FAA = $\ln(P(F4)) - \ln(P(F3))$, where $P$ is EEG power as a function of time in the individual upper alpha EEG band [IAF, IAF+2] Hz for a given channel (F3 or F4). In addition to the FAA, a power-sum function $\ln(P(F4)) + \ln(P(F3))$ was calculated for the upper alpha band. Similarly, the normalized FBA was computed as FBA = $\ln(P(F3)) - \ln(P(F4))$, where $P$ is EEG power as a function of time in the high-beta frequency band [21-30] Hz. A power-sum function was calculated for the same channels for the high-beta band. Average FAA and FBA changes between the Rest and Happy Memories conditions were used to characterize the EEG-nf performance.

*2.10. EEG-based PPI analyses*

To investigate how temporal correlations between FAA (or FBA) and BOLD activity changed between experimental conditions, we performed PPI analyses adapted for EEG-fMRI (Zotev et al., 2014, 2016, 2018a), as described in detail in *Supplementary material* (S1.7). The FAA time course was used to define two PPI regressors: the FAA-based PPI correlation regressor and the FAA-based PPI interaction regressor for the Happy Memories vs Count condition contrast (S1.7, Fig. S3). We chose this contrast a priori as in our previous studies (Zotev et al., 2014, 2016) to compare EEG-fMRI correlations between two cognitive tasks. The FAA-based PPI regressors were orthogonalized with respect to the corresponding PPI regressors based on the EEG power sum, mentioned above. A single-subject EEG-based PPI analysis for each run involved fitting a GLM model with these two PPI regressors (in addition to other fMRI regressors, see S1.7). For FBA, the PPI regressors were defined in a similar way, starting with the FBA time course.

## 3. Results

*3.1. Emotional state changes*

The MDD patients' mood ratings before and after the rtfMRI-EEG-nf session and statistics for the rating changes are reported in Table 1. Five mood ratings most relevant to the present study – POMS depression, confusion, total mood disturbance, STAI state anxiety, and VAS happiness – are included in the table. There were no significant EG vs CG group differences in these ratings before the session. Significant improvements in the mood ratings with medium effect sizes ($d$=−0.62, −0.73, −0.60, −0.57, and +0.59, respectively) were observed after the rtfMRI-EEG-nf session for the EG (Table 1). The corresponding mood improvements for the CG were non-significant with small effect sizes (Table 1). The EG vs CG group differences in the mood rating changes were not significant.

The MDD patients' memory-recall and happiness ratings, reported verbally during the experiment, are included in *Supplementary material* (S2.1, Fig. S4). Both ratings were higher for the EG than for the CG, with the EG vs CG group differences trending toward significance after correction, with large effect sizes ($d$=0.99 and 0.82, respectively, S2.1).

*3.2. Neurofeedback performance*

The main rtfMRI-EEG-nf performance characteristics for the EG participants are exhibited in Figure 4. The results were obtained in offline EEG and fMRI data analyses. To characterize nf performance across the entire nf training, we averaged individual-subject results across the four nf runs (Practice, Run 1, Run 2, Run 3), and compared their group mean to zero using a one-sample $t$-test (two-tailed). The corresponding significance ($p$-value) and effect size (Cohen's $d$) are included at the bottom of each figure after the NF notation.

According to Fig. 4, the EG participants were able to significantly increase the FAA during the rtfMRI-EEG-nf task compared to the Rest condition (NF: $t(15)=3.21$, $p<0.006$), with large effect size ($d=0.80$, 95% CI [0.23 1.36]). They also significantly upregulated the FBA (NF: $t(15)=2.71$, $p<0.016$), with medium effect size ($d=0.68$, 95% CI [0.12 1.21]). Furthermore, the EG participants significantly increased BOLD activity of the LA target

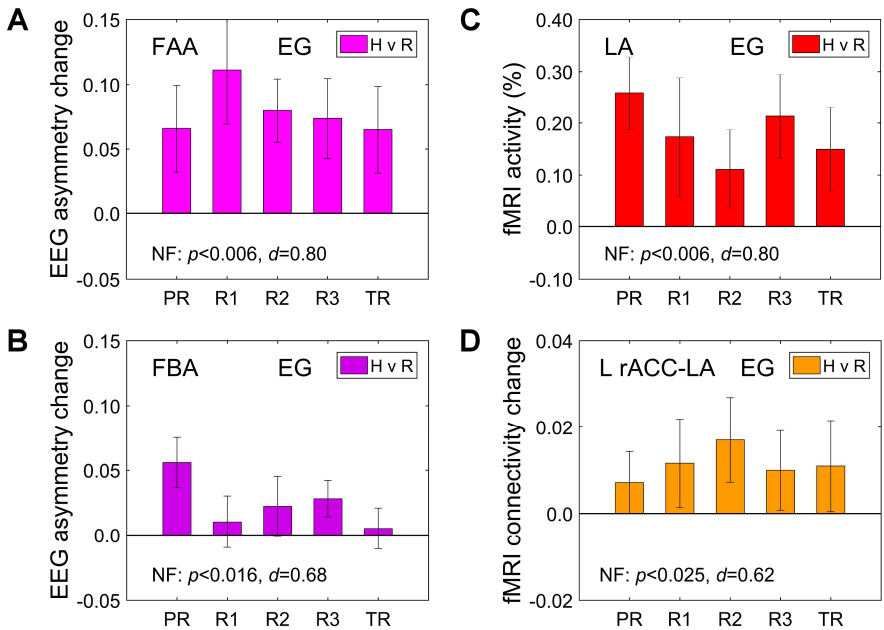

**Figure 4.** Main performance characteristics of the rtfMRI-EEG-nf training for the experimental group (EG). Each bar represents a group mean of average individual results for a given run. The error bars are standard errors of the mean (sem). The experimental runs and condition blocks are depicted schematically in Fig. 1B. The NF at the bottom of each figure refers to group statistics ($p$-value from a $t$-test and effect size $d$, both relative to zero) for the individual results averaged across the four nf runs (PR, R1, R2, R3). A) Average changes in frontal alpha EEG asymmetry (FAA) between the Happy Memories and Rest conditions (H vs R). FAA = $\ln(P(F4)) - \ln(P(F3))$, where $P$ is EEG power in the individual upper alpha band. B) Average changes in frontal high-beta EEG asymmetry (FBA) between the same conditions. FBA = $\ln(P(F3)) - \ln(P(F4))$, where $P$ is EEG power in the high-beta band. C) Average fMRI percent signal changes for the LA target ROI (Fig. 2A) for the Happy Memories conditions with respect to the Rest baseline (H vs R). D) Average changes in fMRI functional connectivity between the LA and L rACC target ROI (Fig. 2B) during the Happy Memories conditions compared to the Rest conditions (H vs R, psychophysiological interaction effect).

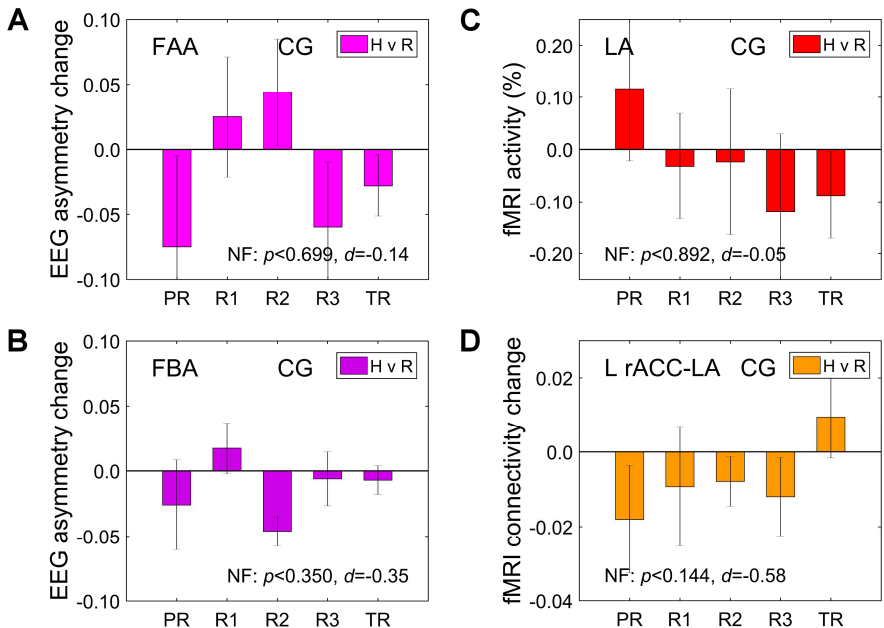

**Figure 5.** Main performance characteristics for the control group (CG). Notations are the same as in Fig. 4.

ROI during the rtfMRI-EEG-nf task relative to the Rest baseline (NF: $t(15)=3.21$, $p<0.006$), with large effect size ($d=0.80$, 95% CI [0.23 1.36]). They also exhibited significant enhancement in fMRI connectivity between the LA and the L rACC target ROI (NF: $t(15)=2.49$, $p<0.025$), with medium effect size ($d=0.62$, 95% CI [0.08 1.15]). The fMRI connectivity changes were computed as the LA-based PPI interaction effects (Sec. 2.8) and averaged within the L rACC ROI.

The average results across the nf runs (NF) in Fig. 4 remained significant after the multiple comparisons correction to account for testing the four quantities (FDR $q<0.012$, 0.021, 0.012, 0.025, respectively). There were no significant differences in these activity measures between the last nf training run (Run 3) and the Transfer run without nf, indicating transfer of the learning effects (TR vs R3: $t(15)=-0.24$, $p<0.817$ for the FAA changes; $t(15)=-1.31$, $p<0.211$ for the FBA changes; $t(15)=-0.76$, $p<0.462$ for the LA activations; $t(15)=0.09$, $p<0.931$ for the L rACC vs LA connectivity changes).

Figure 5 exhibits the corresponding activity measures for the CG. The average results across the four nf runs were non-significant with negative effects (Fig. 5). Importantly, the EG vs CG group differences either were significant or trended toward significance before correction (NF, EG vs CG: $t(22)=2.14$, $p<0.044$, $d=0.93$ for the FAA changes; $t(22)=2.38$, $p<0.027$, $d=1.03$ for the FBA changes; $t(22)=1.84$, $p<0.080$, $d=0.79$ for the LA activations; $t(22)=2.82$, $p<0.010$, $d=1.22$ for the L rACC vs LA connectivity changes). These group differences trended toward significance or remained significant after the multiple comparisons correction (FDR $q<0.059$, 0.054, 0.080, 0.040, respectively).

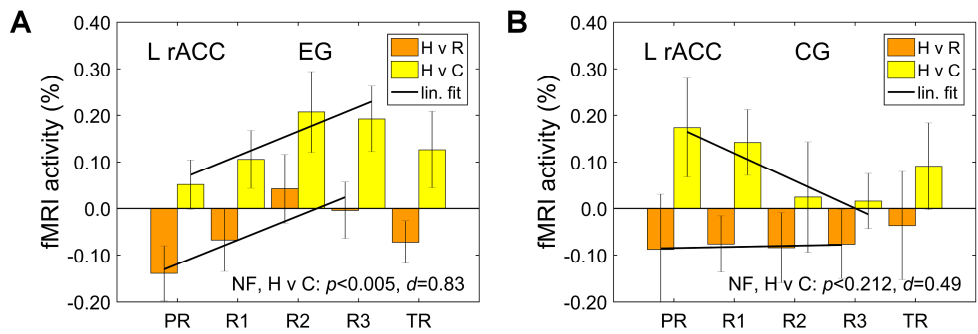

**Figure 6.** Average fMRI percent signal changes for the L rACC target ROI for the Happy Memories vs Rest condition contrast (H vs R) and for the Happy Memories vs Count condition contrast (H vs C). The black straight segments are linear fits across the four nf runs (PR, R1, R2, R3). The NF at the bottom of each figure refers to group statistics (*p*-value from a *t*-test and effect size *d*, both relative to zero) for the individual activity levels, corresponding to the H vs C contrast, averaged across the four nf runs. A) Results for the experimental group (EG). B) Results for the control group (CG).

Figure 6 shows average fMRI percent signal changes for the L rACC target ROI (Fig. 2B) for the Happy Memories vs Rest condition contrast (H vs R) and for the Happy Memories vs Count condition contrast (H vs C) for each group. Statistical analyses for these results are exploratory, because no hypotheses were made about mean L rACC activations. The average L rACC activity levels during the rtfMRI-EEG-nf task across the four nf runs for the EG were non-significant when compared to the Rest baseline (H vs R, NF: $t(15)=-1.13$, $p<0.278$, $d=-0.28$), and were significant when compared to the Count control task (H vs C, NF: $t(15)=3.32$, $p<0.005$, $d=0.83$). The EG vs CG group differences were not significant for either contrast. Interestingly, the L rACC activity levels showed positive linear trends across the four nf runs for both contrasts for the EG (Fig. 6A). We computed a slope of a linear fit to the individual L rACC activity levels across the four nf runs for each participant. For the EG, the mean linear slopes were positive and trended toward significance for both contrasts (H vs R, slope: $t(15)=2.12$, $p<0.051$, $d=0.53$; H vs C, slope: $t(15)=1.75$, $p<0.099$, $d=0.44$). The EG vs CG group difference in the linear slopes was significant, with large effect size, for the Happy Memories vs Count contrast (H vs C, slope difference: $t(22)=2.41$, $p<0.025$, $d=1.04$).

### 3.3. FAA changes vs psychological measures

We conducted exploratory correlation analyses to examine associations between the target activity measures during the rtfMRI-EEG-nf training and individual psychological metrics relevant to MDD. Significant associations were found only for the FAA changes. Results for the EG are exhibited in Figure 7. Correlations with trait measures are shown in Fig. 7A, and correlations with changes in state measures (after vs before the session) are included in Fig. 7B.

The FAA changes during the rtfMRI-EEG-nf task relative to the Rest condition (H vs R), averaged across the four nf runs, showed significant *positive* correlations with the MDD patients' MADRS depression severity ratings ($r=0.52$, $p<0.039$) and SHAPS anhedonia severity ratings ($r=0.71$, $p<0.002$). The corresponding correlation with the HDRS depression severity ratings trended toward significance ($r=0.45$, $p<0.080$). Among the four nf runs, the correlation between the FAA changes and the MADRS depression ratings was most pronounced for the Practice run (PR: $r=0.65$, $p<0.007$), i.e. at the beginning of the nf training.

The same average FAA changes exhibited significant *negative* correlations with the after-vs-before changes in POMS state depression ratings ($r=-0.56$, $p<0.023$) and POMS total mood disturbance ratings ($r=-0.55$, $p<0.028$). The average FAA changes also showed significant negative correlations with changes in POMS confusion ratings ($r=-0.51$, $p<0.044$) and STAI state anxiety ratings ($r=-0.50$, $p<0.049$). Among the four nf runs, the negative correlation between the FAA changes and POMS depression changes was most pronounced for Run 3 (R3: $r=-0.64$, $p<0.008$), i.e. at the end of the nf training.

All the correlation results in Fig. 7 remained significant when they were controlled for the EG participants' age and gender. For the CG, the correlations corresponding to those illustrated in Fig. 7 were not significant.

### 3.4. Prefrontal BOLD laterality

Whole-brain statistical maps of BOLD fMRI activity during the rtfMRI-EEG-nf training for the EG are reported in *Supplementary material* (S2.2, Fig. S5, Table S2). The results demonstrate significant positive Happy Memories vs Count BOLD activity contrast for the left amygdala region and many areas of the limbic system. They also reveal pronounced BOLD laterality for parts of the MidFG and superior frontal gyrus (SFG) (Fig. S5, Table S2).

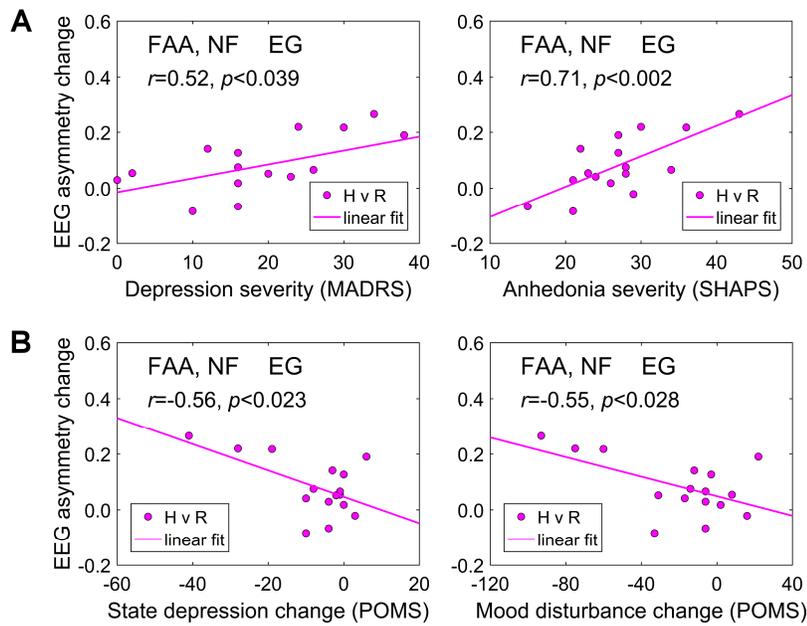

**Figure 7.** Correlations between changes in frontal alpha EEG asymmetry (FAA) during the rtfMRI-EEG-nf training and individual psychological measures. The individual FAA changes between the Rest and Happy Memories with rtfMRI-EEG-nf conditions (H vs R) were averaged across the four nf runs (PR, R1, R2, R3). The results are for the experimental group (EG), with each data point corresponding to one participant. A) Correlations between the FAA changes and severities of depression and anhedonia, assessed before the session. B) Correlations between the FAA changes and changes in state depression and total mood disturbance. The state measures were assessed both before and after the session, and their changes (after vs before) are included in the figures. Acronyms: MADRS – Montgomery-Asberg Depression Rating Scale, SHAPS – Snaith-Hamilton Pleasure Scale, POMS – Profile of Mood States.

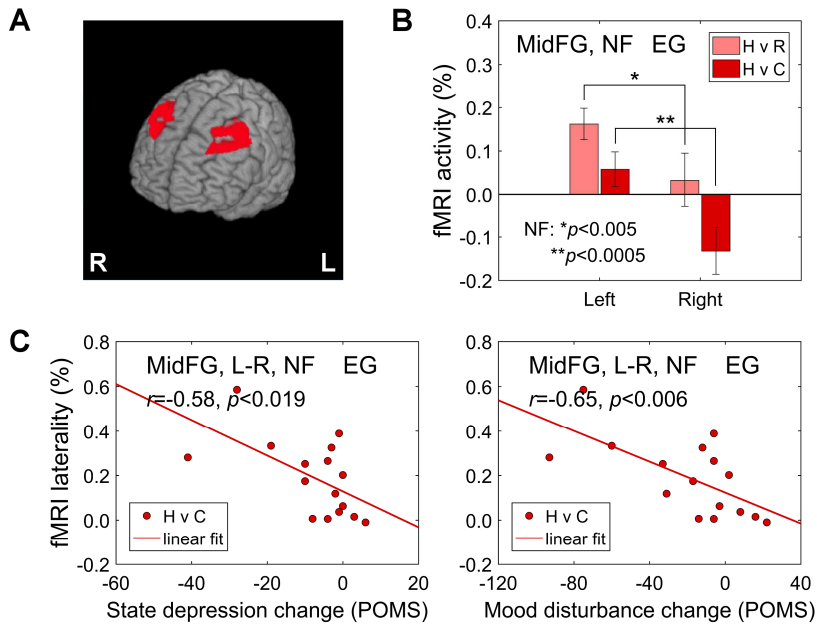

**Figure 8.** Illustration of BOLD laterality for middle frontal gyrus (MidFG) regions during the rtfMRI-EEG-nf training for the experimental group (EG) and its associations with individual psychological measures. A) The left and right MidFG ROIs as seen on the cortical surface. The ROIs were defined as described in the text. B) Average fMRI activity levels (percent signal changes) for the left and right MidFG ROIs, corresponding to the Happy Memories vs Rest condition contrast (H vs R) and to the Happy Memories vs Count contrast (H vs C). The individual fMRI activity levels were averaged across the four nf runs (PR, R1, R2, R3). C) *Left*: Correlation between the average individual MidFG laterality, i.e. the difference in fMRI activities between the left and right MidFG ROIs, for the Happy Memories vs Count contrast and after-vs-before changes in state depression. *Right*: Correlation between the average individual MidFG laterality and changes in total mood disturbance. POMS – Profile of Mood States.

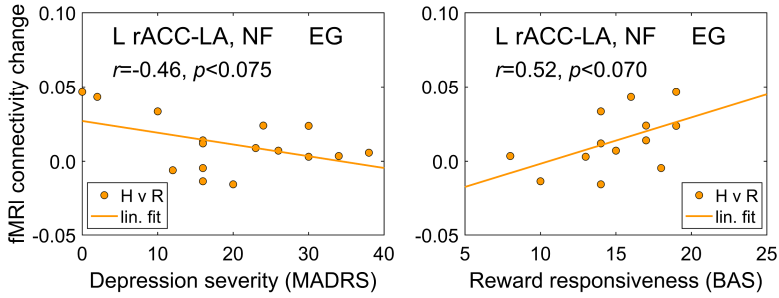

**Figure 9.** Correlations between changes in fMRI functional connectivity of the left amygdala with the left rACC during the rtfMRI-EEG-nf training and individual psychological measures. The fMRI connectivity changes for the target ROIs between the Rest and Happy Memories with rtfMRI-EEG-nf conditions (H vs R, psychophysiological interaction effect) are reported. The results are for the experimental group (EG), with each data point corresponding to one participant (however, $n=13$ for the BAS). The individual results were averaged across the four nf runs (PR, R1, R2, R3). MADRS – Montgomery-Asberg Depression Rating Scale, BAS – Behavioral Activation System scale.

We performed exploratory analyses of prefrontal BOLD laterality to evaluate effects of the EEG-nf, targeting the FAA and FBA, independently of the EEG data analysis. Results for the EG are exhibited in Figure 8. Because EEG electrodes F3 and F4 are situated above the MidFG, we considered the left and right MidFG ROIs, defined as described in *Supplementary material* (S1.4). Following the laterality pattern in Fig. S5, we limited the ROIs to $42 \leq z \leq 57$ mm. The resulting left and right MidFG ROIs are depicted in Fig. 8A. Mean BOLD activity levels for these ROIs, with individual results averaged across the four nf runs, are shown in Fig. 8B. These activity levels were significantly higher for the left MidFG ROI than for the right MidFG ROI (based on paired *t*-test) both for the Happy Memories vs Rest condition contrast (H vs R: $t(15)=3.28$, $p<0.005$, $d=0.82$) and for the Happy Memories vs Count contrast (H vs C: $t(15)=4.46$, $p<0.0005$, $d=1.11$).

We also conducted exploratory correlation analyses to evaluate associations between the MidFG BOLD laterality during the rtfMRI-EEG-nf training and mood rating changes. Results for the EG are shown in Fig. 8C. The average MidFG BOLD laterality for the Happy Memories vs Count contrast exhibited significant negative correlations with the after-vs-before changes in POMS state depression ratings ($r=-0.58$, $p<0.019$) and POMS total mood disturbance ratings ($r=-0.65$, $p<0.006$). For the Happy Memories vs Rest contrast, the laterality also showed negative correlations with changes in these ratings (H vs R, POMS state depression change: $r=-0.48$, $p<0.061$; POMS total mood disturbance change: $r=-0.57$, $p<0.020$). For the CG, the results corresponding to those in Figs. 8B,C were not significant.

### 3.5. Amygdala-rACC connectivity enhancement

Whole-brain statistical maps for the EG vs CG group difference in the LA fMRI connectivity changes during the rtfMRI-EEG-nf training are reported in *Supplementary material* (S2.3, Fig. S6, Table S3). The maps revealed three loci in the rACC area, characterized by the most pronounced EG vs CG group differences (S2.3): (−8, 34, 7), (−9, 41, 5), and (3, 35, 9). All three loci were in close proximity to the center of the L rACC target ROI at (−3, 34, 5). For 10-mm-diameter ROIs, centered at these loci, the EG vs CG group differences in fMRI connectivity changes with the LA were significant with large effect sizes (S2.3).

We performed exploratory correlation analyses to examine associations between the average changes in fMRI connectivity between the LA and the L rACC target ROI during the rtfMRI-EEG-nf training and individual psychological measures relevant to MDD. Figure 9 shows such associations for the EG. The fMRI connectivity changes for the Happy Memories condition relative to the Rest condition (H vs R) were averaged across the four nf runs. These changes exhibited negative and trending toward significance correlation with the MADRS depression severity ratings ($r=-0.46$, $p<0.075$) and positive correlation with the BAS reward responsiveness ratings ($r=0.52$, $p<0.070$). For the CG, the corresponding correlations were not significant.

### 3.6. EEG-fMRI correlations for the left amygdala

Figure 10 reports changes in temporal correlations between time courses of the FAA and FBA and the LA BOLD activity during the rtfMRI-EEG-nf training. The correlation changes were computed as FAA-based or FBA-based PPI interaction effects (Sec. 2.10) and averaged within the LA ROI.

Fig. 10A demonstrates that temporal correlation between the FAA and the LA BOLD activity was significantly enhanced during the rtfMRI-EEG-nf task compared to the Count condition for the EG (NF: $t(15)=2.81$, $p<0.013$), with medium effect size ($d=0.70$, 95% CI [0.14 1.24]). There was no significant difference in the mean PPI interaction effects between Run 3 and the Transfer run (TR vs R3: $t(15)=0.22$, $p<0.833$). Similarly, Fig. 10B indicates significant enhancement in temporal correlation between the FBA and the LA BOLD activity during the rtfMRI-EEG-nf task for the EG (NF: $t(15)=2.51$, $p<0.024$), with medium effect size ($d=0.63$, 95% CI [0.08 1.16]). However, the PPI effect was negligible during the Transfer run in this case. For the CG, the FAA-based PPI interaction effect was negative yet small (Fig. 10C), while the FBA-based PPI interaction effect was negative with large effect size (Fig. 10D).

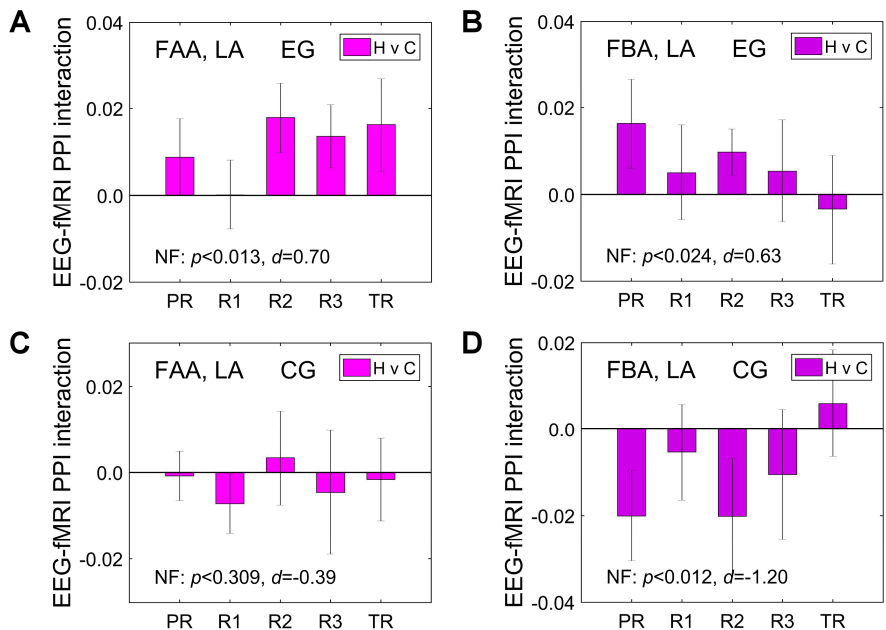

**Figure 10.** Changes in temporal correlations between frontal EEG asymmetries and BOLD fMRI activity of the left amygdala during the rtfMRI-EEG-nf training. The correlation changes between Happy Memories and Count conditions (H vs C) were evaluated in EEG-fMRI psychophysiological interaction (PPI) analyses, based on time courses of frontal alpha EEG asymmetry (FAA) and frontal high-beta EEG asymmetry (FBA). Each bar represents a group mean of the individual PPI interaction values for a given run, averaged within the LA ROI. The error bars are standard errors of the mean (sem). The NF at the bottom of each figure refers to group statistics ($p$-value from a $t$-test and effect size $d$, both relative to zero) for the individual results averaged across the four nf runs (PR, R1, R2, R3). A) Average values of the FAA-based PPI interaction effect for the LA ROI for the experimental group (EG). B) Average values of the FBA-based PPI interaction effect for the LA ROI for the EG. C) Average values of the FAA-based PPI interaction effect for the control group (CG). D) Average values of the FBA-based PPI interaction effect for the CG.

Importantly, the EG vs CG group differences in the PPI interaction effects for the LA ROI were significant with large effect sizes for both the FAA ($t(22)=2.32$, $p<0.030$, $d=1.00$) and the FBA ($t(22)=3.90$, $p<0.001$, $d=1.69$).

Average individual values of the FAA-based PPI interaction effect for the LA ROI across the four nf runs for the EG exhibited significant positive correlation with the corresponding average values of the amygdala BOLD laterality (NF: $r=0.53$, $p<0.035$), as illustrated in *Supplementary* Fig. S7A. When the individual results were averaged across three nf runs (out of four) characterized by the most positive amygdala BOLD laterality values, the correlation was more significant (NF*: $r=0.61$, $p<0.012$), as shown in Fig. S7B. The amygdala BOLD laterality ('LA−RA') was computed as described in *Supplementary material* (S1.4). It can be viewed as a performance measure characterizing *target-specific* effects of the LA-based rtfMRI-nf procedure, as explained in *Supplementary material* (S2.4, see also Zotev et al., 2016).

Consistent with the results for the LA in Fig. 10A, the average FAA-based PPI interaction effect for the L rACC target ROI was positive and trended toward significance for the EG (NF: $t(15)=2.03$, $p<0.061$, $d=0.51$). The EG vs CG group difference in the FAA-based PPI interaction effects for the L rACC ROI also trended toward significance ($t(22)=1.86$, $p<0.076$, $d=0.81$).

### 3.7. EEG-fMRI correlations across the brain

Figure 11 exhibits whole-brain statistical maps for the FAA-based PPI interaction effect corresponding to the Happy Memories vs Count condition contrast for the EG. The PPI interaction results from twelve EG participants were included in the group analysis. The other four cases were considered outliers based on the low amygdala BOLD laterality (Fig. S7). For each participant, the PPI interaction maps were averaged for three nf runs (out of four) with the most positive individual amygdala BOLD laterality values. Statistical results for the FAA-based PPI interaction effect are summarized in Table 2. The maps in Fig. 11 are FDR corrected with $q<0.04$ threshold, and the data in Table 2 – with $q<0.02$ threshold. Whole-brain statistical maps for the FBA-based PPI interaction effect, obtained in a similar way for the same contrast for the EG, are reported in *Supplementary material* (S2.5, Fig. S8, Table S4).

Figure 12 compares the FAA- and FBA-based PPI interaction effects from Fig. 11 and Fig. S8 for the left amygdala and its vicinity. The PPI results demonstrate that both the FAA and FBA exhibited enhanced temporal correlations, during the rtfMRI-EEG-nf task, with BOLD activities of the left amygdala and large brain networks. These results are discussed in detail below.

## 4. Discussion

In this paper, we reported the first, proof-of-concept application of the simultaneous real-time fMRI and EEG neurofeedback (rtfMRI-EEG-nf) for emotion self-regulation training in patients with a neuropsychiatric disorder, specifically, major depressive disorder (MDD). This is also the first neurofeedback study in which participants had an opportunity to simultaneously regulate two rtfMRI-nf signals and two EEG-nf signals.

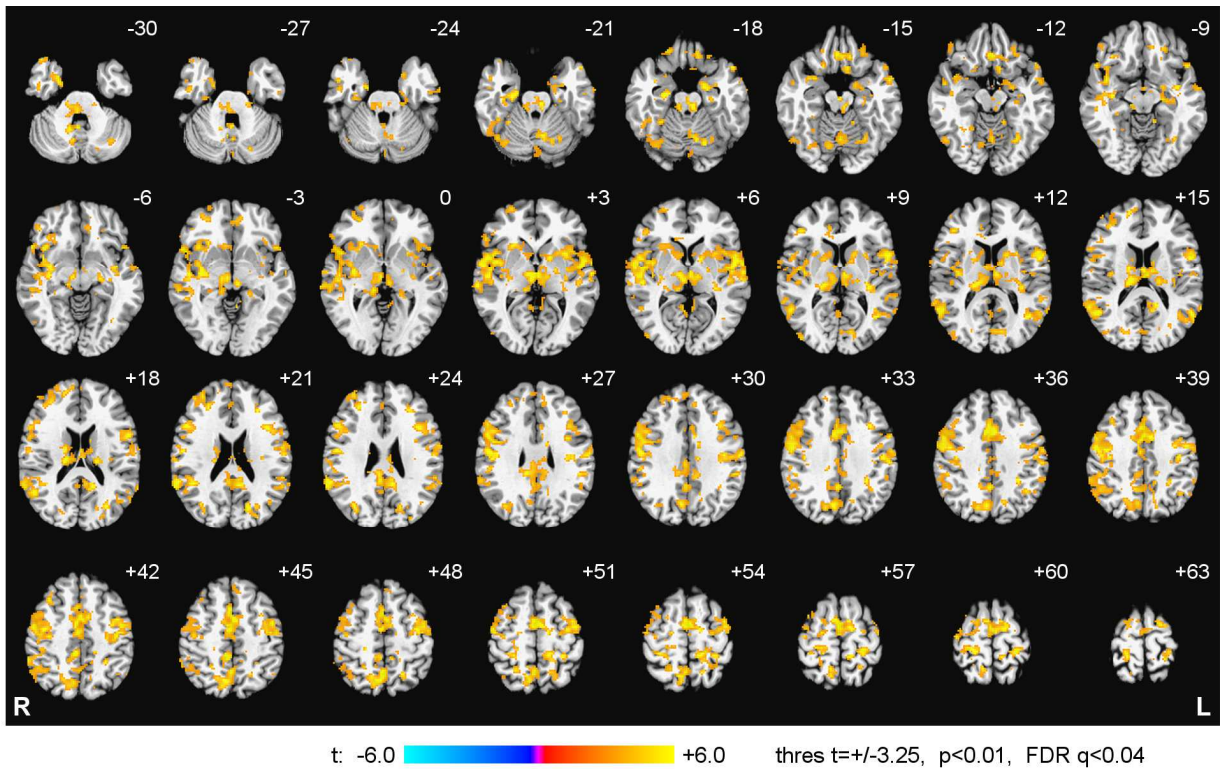

**Figure 11.** Enhancement in temporal correlation between frontal alpha EEG asymmetry (FAA) and BOLD activity during the rtfMRI-EEG-nf training. Statistical maps of the FAA-based PPI interaction effect for the Happy Memories vs Count condition contrast (H vs C) are shown for the experimental group (EG). The maps are voxel-wise FDR corrected and projected onto the standard TT_N27 anatomical template, with 3 mm separation between axial slices. The number adjacent to each slice indicates the $z$ coordinate in mm. The left hemisphere (L) is to the reader's right. Peak $t$-statistics values for the FAA-based PPI interaction effect and the corresponding locations are specified in Table 2.

Furthermore, we implemented the advanced real-time EEG-fMRI artifact correction procedure that made EEG-nf during fMRI practical and efficient.

### 4.1. Emotional state changes

Following the rtfMRI-EEG-nf session, the MDD patients in the EG showed significant mood improvements, including significant reductions in state depression, confusion, total mood disturbance, and state anxiety, as well as significant increase in state happiness (Table 1). These improvements, characterized by medium effect sizes, suggest that the rtfMRI-EEG-nf training may be beneficial to MDD patients. The significant reduction in confusion may indicate that the EG participants were able to develop a better grasp of the experimental procedure as the training continued, despite the relative complexity of the rtfMRI-EEG-nf task. During the experiment, the EG participants rated their abilities to recall autobiographical memories and feel happiness higher, than the CG participants (*Supplementary* Fig. S4). These group differences, with large effect sizes, suggest that the rtfMRI-EEG-nf provided information more consistent with the Happy Memories task than the sham feedback.

### 4.2. Neurofeedback performance

During the rtfMRI-EEG-nf training, the MDD patients in the EG learned to significantly increase BOLD activity of the LA (Fig. 4C) and significantly upregulate the FAA (Figs. 4A). These are the primary outcome measures in our study. The EG participants also achieved significant enhancement in fMRI connectivity between the LA and the L rACC (Fig. 4D) and significant upregulation of the FBA (Fig. 4B), which are the secondary outcome measures. These results support the first of the two main hypotheses in our study (Sec. 1). Importantly, the EG vs CG group differences in these four measures either were significant or trended toward significance (Sec. 3.2), indicating that effects of the rtfMRI-EEG-nf were specific to the EG and different from those of the sham feedback for the CG. The left rACC activity levels exhibited positive linear trends across the four nf runs for the EG (Fig. 6).

In our previous study, which combined rtfMRI-nf of the LA activity with passive EEG (Zotev et al., 2016), the effect size for the LA activation was large ($d$=0.87, 95% CI [0.21 1.50]), while the effect size for the associated FAA increase was small ($d$=0.45, 95% CI [−0.14 1.01]). In the present work, both effects sizes were large ($d$=0.80, 95% CI [0.23 1.36], Figs. 4A,C). This comparison (which should be taken with caution because of the wide confidence intervals) suggests potential benefits of the rtfMRI-EEG-nf compared to the rtfMRI-nf alone.

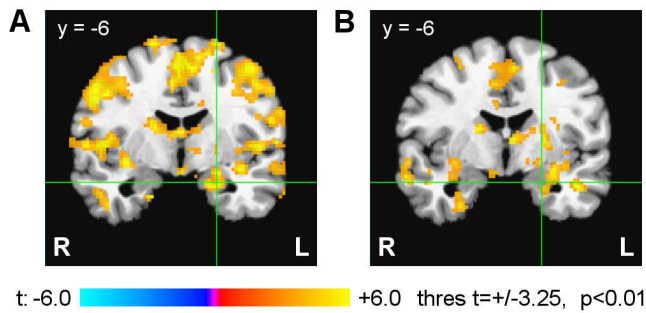

**Figure 12.** Comparison of the FAA- and FBA-based PPI interaction effects for the left amygdala region. A) FAA-based PPI interaction effects from Fig. 11. B) FBA-based PPI interaction effects from Supplementary Fig. S8. The green crosshairs ($x$=−21 mm, $z$=−16 mm) correspond to the center of the LA target ROI.

*4.3. FAA changes vs psychological measures*

The average individual FAA changes during the rtfMRI-EEG-nf task for the EG showed significant *positive* correlations with the MDD patients' depression and anhedonia severities (Fig. 7A). These findings are consistent with those reported in our study of EEG correlates of the amygdala rtfMRI-nf (Zotev et al., 2016). MDD patients exhibit lower FAA levels compared to non-depressed individuals, particularly during an emotional challenge (Stewart et al., 2011, 2014). The more positive FAA changes during the rtfMRI-EEG-nf task in the patients with more severe depression suggest the potential for correction of the FAA deficiencies specific to MDD (Zotev et al., 2016). This reasoning is explained in more detail in the follow-up study (Zotev and Bodurka, 2020). It is supported by the observed mood improvements: the MDD patients, who achieved more positive average FAA changes during the rtfMRI-EEG-nf task, showed stronger reductions in state depression and total mood disturbance after the training (Fig. 7B).

The positive correlations between the FAA changes and the depression and anhedonia severities in Fig. 7A are not as pronounced as those in our previous work (Zotev et al., 2016). The primary reason is that the FAA in the present study was explicitly modulated via the EEG-nf. Consequently, the FAA changes depended on individual EEG-nf performance, including a participant's attention to the FAA-based EEG-nf signal and effort to regulate it. Indeed, the positive correlation between the FAA changes and the depression severity was most pronounced for the Practice run, when the participants were first exposed to the rtfMRI-EEG-nf, and became weaker as the training continued (Sec. 3.3).

*4.4. Prefrontal BOLD laterality*

Performance of the rtfMRI-EEG-nf task was associated with pronounced laterality of BOLD activations for the dorsal PFC regions (*Supplementary* Fig. S5). BOLD activity levels during the rtfMRI-EEG-nf task (relative to the Count condition) were more positive for the MidFG and SFG areas on the left, compared to the corresponding MidFG and SFG areas on the right (Fig. S5). The most positive BOLD contrast *t*-score was observed for the left MidFG (BA 8) at (−40, 20, 47), while the most negative contrast *t*-score occurred in the right SFG (BA 8) at (27, 17, 49) (Table S2). These locations are parts of the left and right DLPFC, respectively. Furthermore, EEG electrodes F3 and F4 are situated above the MidFG and BA 8. Thus, the significant positive MidFG BOLD laterality, illustrated in Fig. 8B, is consistent with the significant positive FAA and FBA changes for channels F3 and F4 (Figs. 4A,B). The negative correlations between the average MidFG BOLD laterality values and the after-vs-before changes in the mood ratings (Fig. 8C, Sec. 3.4) are consistent with the negative correlations between the average FAA changes and the same mood rating changes (Fig. 7B). Therefore, the prefrontal BOLD laterality effects independently confirm the EEG asymmetry effects observed in the EEG data analyses.

*4.5. Amygdala-rACC connectivity enhancement*

Modulation of the left rACC BOLD activity simultaneously with that of the LA (Fig. 1A) enabled (or, at least, was consistent with) enhancement in the two regions' fMRI functional connectivity (Fig. 4D). This effect was specific to the rtfMRI-EEG-nf, as evidenced by the significant EG vs CG group differences in fMRI connectivities of the LA with three loci close to the center of the L rACC target ROI (*Supplementary* Fig. S6B). In future studies, an rtfMRI-nf based on an actual fMRI connectivity metric, such as the Pearson's correlation coefficient, could be implemented and used together with the rtfMRI-nf of the amygdala activity. The enhancement in fMRI connectivity between the LA and the left rACC during the rtfMRI-EEG-nf task for the EG showed negative association with the MDD patients' depression severity and positive association with reward responsiveness (Fig. 9). This means that a stronger interaction between these two regions during positive emotion induction with rtfMRI-EEG-nf should be beneficial to MDD patients. This observation is consistent with results of the previous studies that emphasized the important role of the rACC in emotion regulation and modulation on the amygdala activity (e.g. Etkin et al., 2006; Pizzagalli, 2011; Zotev et al., 2013).

*4.6. EEG-fMRI correlations for the left amygdala*

The FAA-based PPI interaction results for the LA ROI (Fig. 10A) demonstrate that temporal correlation between the FAA time course (convolved with the HRF) and the LA BOLD activity was significantly stronger during the rtfMRI-EEG-nf task than during the control condition. Similarly, the FBA-based PPI interaction results (Fig. 10B) indicate significant enhancement in temporal correlation between the FBA time course and the LA BOLD activity. These EEG-based PPI interactions had medium effect sizes for the EG, and large effect sizes when compared to those for the CG (Sec. 3.6). Therefore, the EG participants were able, on the average, to upregulate the FAA and FBA together with the LA activity. These results support the second of the two main hypotheses in our study (Sec. 1).

Interestingly, the average individual FAA-based PPI interaction effects for the LA showed significant positive correlation with the average individual amygdala BOLD laterality (*Supplementary material* S2.4, Fig. S7). This means that the EG participants, who were more successful at upregulating BOLD activity of the target amygdala region (LA) relative to the non-target region (RA), were also more successful at doing so simultaneously with increasing the FAA-based EEG-nf signal. This observation confirms the connection between the FAA changes and the amygdala BOLD laterality we reported previously (Zotev et al. 2016).

*4.7. fMRI correlates of the FAA modulation*

The whole-brain maps of the FAA-based PPI interaction effect (Fig. 11) demonstrate that temporal correlation between the FAA and BOLD activity was significantly enhanced, during the rtfMRI-EEG-nf task for the EG, not only for the left amygdala, but also for the large brain network. Note that the FAA- and FBA-based PPI regressors were orthogonalized with respect to the corresponding regressors based on the EEG power sums for channels F3 and F4. Therefore, a positive PPI interaction effect for a given region means that its BOLD activity increased simultaneously with activation of cortical areas contributing to EEG signal measured by F3, and decreased simultaneously with deactivation of areas contributing to EEG signal measured by F4.

The results in Fig. 11 show significant positive PPI interaction effects for the corresponding left and right DLPFC regions (MidFG, BA 9), with maxima at (−36, 16, 24) and (49, 18, 24), respectively (Fig. 11, Table 2). These results are consistent with the common view of frontal EEG asymmetry as reflecting activation of the left DLPFC and deactivation of the right DLPFC (and vice versa). Within the approach-avoidance framework, these effects are interpreted as indicative of enhanced approach motivation and reduced avoidance motivation, respectively.

The results in our study demonstrate involvement of the left premotor cortex (PMC), specifically the precentral gyrus (PrecG), BA 6 in performance of the rtfMRI-EEG-nf task. A local maximum of the FAA-based PPI interaction effect is observed near the border of the left PrecG and MidFG at (−42, −4, 44) (Fig. 11, Table 2). This locus is relatively close to the location of the maximum of the corresponding BOLD activity contrast in the left MidFG at (−40, 20, 47) (Fig. S5, Table S2), which is also near the anterior boundary of the PrecG. Furthermore, the EG vs CG group difference in the LA connectivity enhancement during the rtfMRI-EEG-nf task was prominent in nearby regions of the left MidFG at (−53, 8, 36) and PrecG at (−51, 1, 33) (Fig. S6A, Table S3). Collectively, these findings point to mutually consistent roles of the left DLPFC and the adjacent area of the left PMC during the rtfMRI-EEG-nf task.

The last observation is not surprising, because the anterior (rostral) PMC has strong interconnections with the prefrontal cortex (e.g. Chouinard and Paus, 2006; Hanakawa et al., 2003). A recent meta-analysis of rtfMRI-nf studies with various target regions revealed consistent fMRI activations of bilateral DLPFC areas extending to PMC (Emmert et al., 2016). We hypothesize that the left DLPFC in our study is involved in mental strategy implementation, while the left PMC is involved in observation and control of the variable-height nf bars. From this point of view, the enhanced temporal correlation between the FAA and the left PMC activity (Fig. 11, Table 2) suggests that the FAA modulation was closely associated with direct regulation of the FAA-based EEG-nf signal.

Interestingly, resting-state EEG source imaging studies have suggested that motivation is related to activities of both the DLPFC and the PMC. Stronger reward bias in healthy individuals is associated with reduced upper alpha (alpha2) EEG activity (i.e. stronger activation) in the left MidFG, left SFG, and left PrecG (BA 6) (Pizzagalli et al., 2005). In MDD patients, resting alpha EEG source laterality index shows negative correlations with depression severity for both the MidFG and the PrecG regions (Smith et al., 2018). It is suggested that the PMC activity may "facilitate mobilization of the body for approach-motivated behaviors" (Smith et al., 2018). In the hierarchical model of approach/avoidance motivation by Sprielberg et al., 2013, the left DLPFC instantiates approach motivation at the strategic level, while the left PMC subserves it at the tactical level.

In the left amygdala area, the main statistical maximum for the FAA-based PPI interaction effect is observed in the superficial (SF) subdivision of the amygdala at (−17, −5, −17) (Fig. 12A, Table 2). This finding is consistent with that in our previous study (Zotev et al., 2016), which showed that the same PPI effect had the maximum in the SF subdivision at (−17, −3, −16) (Table 2 therein). Two additional maxima are found in the laterobasal (LB) amygdala subdivision at (−21, −6, −19) and (−21, −6, −11) (Fig. 12A, Table 2). Compared to the LB, the SF subdivision is more closely involved in processing reward-related and socially relevant information, as well as in modulation of approach-avoidance behavior (Bzdok et al., 2013).

*4.8. fMRI correlates of the FBA modulation*

The FBA-based PPI interaction effects are most pronounced along the cortical midline and the cingulate gyrus (*Supplementary material* S2.5, Fig. S8, Table S4). Elevated high-beta activity in these areas, often with some lateralization to the right, is associated with anxiety (e.g. Zotev and Bodurka, 2020, and references therein). Significant FBA-based PPI interaction effects are also found for many regions involved in autobiographical memory retrieval, including the hippocampus, the extended areas of the parahippocampal gyrus, the anterior thalamus, the precuneus, the posterior cingulate, the lingual gyrus (involved in visual memory), and others (Fig. S8, Table S4). For the PMC areas, the PPI effects are less pronounced than those for the FAA. These findings suggest that the FBA modulation during the rtfMRI-EEG-nf task might have been more closely associated with variations in anxiety and activity of the autobiographical memory system than with direct regulation of the FBA-based EEG-nf signal. Indeed, high-beta EEG activity is relevant to the autobiographical memory function and limbic functions in general (e.g. Cannon et al., 2005; Paquette et al., 2009).

In the left amygdala region, the main statistical maximum for the FBA-based PPI interaction effect is observed in the LB amygdala subdivision at (−28, −5, −10) (Fig. 12B, Table S4). This result is consistent with that in our previous work (Zotev et al., 2014), which showed that the same PPI effect was more pronounced in the LB subdivision of the left amygdala (Fig. 4 therein). Therefore, while the FAA temporal variations during the rtfMRI-EEG-nf task exhibited enhanced correlations with BOLD activities of both the SF and LB amygdala subdivisions, the FBA variations showed enhanced correlation mainly with activity of the LB subdivision.

*4.9. Study limitations*

The reported study has several limitations. First, the experimental protocol with happy emotion induction based on recall of happy autobiographical memories was adopted from our earlier studies on the amygdala rtfMRI-nf (Zotev et al., 2011; also Young et al., 2014), and was not optimized for the rtfMRI-EEG-nf. In future studies, mental strategies and training procedures will have to be developed specifically for the rtfMRI-EEG-nf to maximize simultaneous engagement of both fMRI and EEG target brain activities. Second, the study participants had, on the average, moderate depression (Table S1). Recruitment of more unmedicated MDD patients with severe depression will help to elucidate effects of the rtfMRI-EEG-nf that are specific to MDD. Third, the sham feedback signals, provided to the control group participants, were computer generated and unrelated to brain activity. Further research on rtfMRI-EEG-nf will benefit from a more realistic sham feedback, utilizing actual real-time fMRI and EEG data.

## 5. Conclusion

Our simultaneous real-time fMRI and EEG neurofeedback (rtfMRI-EEG-nf) procedure provided proof-of-concept demonstration of intended target engagements and modulatory effects on recruited brain circuitry dynamics. Furthermore, we observed enhanced temporal correlations of the target EEG and fMRI activity measures, clearly indicating the ability of both neurofeedback modalities to capture common aspects of neuronal activity. In our opinion, the rtfMRI-EEG-nf is worth implementation efforts, because it is a powerful approach to influence brain activity in a more experimentally controllable fashion and investigate resulting changes in spatial and temporal brain dynamics. Our study suggests that the rtfMRI-EEG-nf can benefit depressed individuals and may have potential for treatment of MDD. The described rtfMRI-EEG-nf implementation with two rtfMRI-nf and two EEG-nf signals is an advanced and versatile neuromodulation tool. Efficient mental strategies and imaginative experimental designs will be needed to take full advantage of the opportunities it offers. Ultimately, effectiveness of the rtfMRI-EEG-nf compared to either of the individual neurofeedback modalities will have to be demonstrated.

**Conflict of interest**

The authors declare that the research was conducted in the absence of any commercial or financial relationships that could be construed as a potential conflict of interest.

**Funding**

This work was supported by the Laureate Institute for Brain Research and the William K. Warren Foundation, and in part by the P20 GM121312 award from National Institute of General Medical Sciences, National Institutes of Health.

**Acknowledgments**

We would like to thank Dr. Tracy Warbrick and Dr. Brett Bays of Brain Products, GmbH for their continued help, inspired teaching, and excellent technical support.

**Table 1.** Participants' emotional state measures before and after the rtfMRI-EEG-nf session. Emotional states were assessed using the Profile of Mood States (POMS), the State-Trait Anxiety Inventory (STAI), and the Visual Analogue Scale (VAS).

| Measure | Before mean (SD) | After mean (SD) | Effect size ($d$) | Change $t$-score# | Change $p$-value [$q$] |
|---|---|---|---|---|---|
| **Experimental group (EG, $n$=16)** | | | | | |
| POMS | | | | | |
|   Depression | 15.4 (14.0) | 7.75 (10.1) | −0.62 | −2.49 | 0.025 [0.039]* |
|   Confusion | 10.7 (5.91) | 7.25 (4.22) | −0.73 | −2.91 | 0.011 [0.039]* |
|   Total mood disturbance | 46.1 (39.9) | 26.8 (28.0) | −0.60 | −2.39 | 0.030 [0.039]* |
| STAI | | | | | |
|   State anxiety | 44.9 (12.2) | 40.4 (10.5) | −0.57 | −2.26 | 0.039 [0.039]* |
| VAS | | | | | |
|   Happiness | 4.56 (2.31) | 5.94 (1.53) | +0.59 | +2.36 | 0.033 [0.039]* |
| **Control group (CG, $n$=8)** | | | | | |
| POMS | | | | | |
|   Depression | 23.9 (17.4) | 19.6 (13.4) | −0.42 | −1.18 | 0.276 [0.856] |
|   Confusion | 11.4 (5.21) | 10.4 (4.21) | −0.29 | −0.83 | 0.436 [0.856] |
|   Total mood disturbance | 65.8 (40.4) | 59.0 (34.8) | −0.21 | −0.58 | 0.580 [0.856] |
| STAI | | | | | |
|   State anxiety | 53.8 (12.0) | 53.3 (9.51) | −0.04 | −0.11 | 0.915 [0.915] |
| VAS | | | | | |
|   Happiness | 2.38 (1.85) | 2.75 (2.25) | +0.15 | +0.42 | 0.685 [0.856] |

# $t(15)$ for the EG, $t(7)$ for the CG, two-tailed. * FDR $q<0.05$ for the five tests.

**Table 2.** Psychophysiological interaction effect, based on the time course of frontal alpha EEG asymmetry (FAA), for the Happy vs Count condition contrast for the experimental group (EG).

| Region | Laterality | x, y, z (mm) | t-score |
|---|---|---|---|
| **Frontal lobe** | | | |
| Medial frontal gyrus (BA 6) | L | −3, −5, 54 | 17.6 |
| Superior frontal gyrus (BA 6) | L | −21, −5, 65 | 12.1 |
| Precentral gyrus (BA 4) | L | −39, −15, 42 | 11.5 |
| Precentral gyrus (BA 6) | R | 49, −11, 26 | 10.3 |
| Medial frontal gyrus (BA 11) | R | 2, 29, −14 | 9.51 |
| Inferior frontal gyrus (BA 46) | R | 39, 33, 10 | 9.28 |
| Inferior frontal gyrus (BA 44) | L | −51, 10, 12 | 8.89 |
| Precentral gyrus (BA 4) | R | 45, −11, 51 | 8.86 |
| Inferior frontal gyrus (BA 47) | R | 37, 16, −4 | 8.71 |
| Middle frontal gyrus (BA 9) | R | 49, 18, 24 | 8.23 |
| Precentral / mid. frontal gyrus (BA 6) | L | −42, −4, 44 | 7.81 |
| Middle frontal gyrus (BA 9) | L | −36, 16, 24 | 6.33 |
| **Temporal lobe** | | | |
| Superior temporal gyrus (BA 22) | L | −51, 1, 5 | 11.1 |
| Superior temporal gyrus (BA 38) | L | −47, 15, −18 | 8.73 |
| Superior temporal gyrus (BA 39) | R | 53, −53, 12 | 8.53 |
| Middle temporal gyrus (BA 19/22) | L | −38, −59, 12 | 8.34 |
| Superior temporal gyrus (BA 22) | R | 50, −15, 6 | 7.90 |
| Superior temporal gyrus (BA 39) | L | −51, −55, 14 | 7.02 |
| Transverse temporal gyrus (BA 41) | L | −29, −29, 8 | 6.95 |
| **Parietal lobe** | | | |
| Inferior parietal lobule (BA 40) | R | 59, −43, 24 | 12.6 |
| Precuneus (BA 7) | R | 5, −35, 44 | 10.2 |
| Precuneus (BA 31/18) | L | −17, −68, 21 | 10.1 |
| Precuneus (BA 7) | R | 1, −55, 48 | 9.57 |
| Postcentral gyrus (BA 4) | R | 19, −33, 62 | 8.49 |
| Postcentral gyrus (BA 3) | L | −27, −31, 60 | 8.40 |
| Precuneus (BA 31) | R | 1, −48, 33 | 7.90 |
| **Limbic lobe** | | | |
| Cingulate gyrus (BA 24) | L | −1, 2, 44 | 13.1 |
| Amygdala / parahipp. gyrus | L | −17, −5, −17 | 8.73 |
| Parahippocampal gyrus (BA 28) | R | 21, −13, −18 | 8.70 |
| Amygdala / uncus | L | −21, −6, −19 | 7.86 |
| Posterior cingulate (BA 29) | L | −12, −47, 18 | 7.85 |
| Amygdala / parahipp. gyrus | L | −21, −6, −11 | 7.73 |
| Uncus (BA 36) | R | 21, −3, −32 | 7.53 |
| Parahippocampal gyrus (BA 28) | L | −21, −19, −8 | 6.94 |
| Subcallosal gyrus (BA 25) | L | −9, 15, −14 | 6.72 |
| **Sub-lobar** | | | |
| Thalamus, laterodorsal | R | 13, −19, 16 | 9.49 |
| Claustrum | R | 33, −7, −4 | 9.38 |
| Insula (BA 13) | R | 44, 0, 4 | 9.12 |
| Thalamus, mediodorsal | L | −3, −13, 6 | 9.10 |
| Thalamus, mediodorsal | R | 3, −15, 2 | 8.80 |
| Culmen | L | −1, −57, −22 | 8.58 |
| Declive | L | −19, −65, −18 | 7.47 |
| Caudate body | R | 17, −9, 20 | 6.82 |

FDR $q<0.02$ for $|t|>6.0$; BA – Brodmann areas; L – left; R – right; x, y, z – Talairach coordinates.

# Supplementary material

## S1.1. Modified MR-compatible EEG cap

To achieve more efficient real-time EEG-fMRI artifact suppression, we modified a 32-channel BrainCap-MR (EASYCAP, GmbH) as shown in Fig. 2D. In the modified cap, four EEG channels out of 31 – FC1, FC2, TP9, TP10 – were re-purposed for acquisition of reference artifact waveforms, which we refer to as $R_1(t)$, $R_2(t)$, $R_3(t)$, and $R_4(t)$, instead of EEG activity. For each of these channels, the lead was disconnected from its electrode, and connected to one end of a wire contour. The other end of the contour was connected to the Ref electrode (FCz, blue) via a 50 kOhm resistor. Geometries of the four contours were optimized so that electromotive forces (EMFs), induced in the contours during head movements in the MRI scanner's main field, approximate cardioballistic (CB) and random-motion artifacts picked up by EEG channels F3 and F4. The two shorter contours (brown wires in Fig. 2D) followed the leads of channels F3 and F4, respectively, then looped around the Gnd electrode (AFz, black), and connected to the Ref via the resistors. The two longer contours (orange wires in Fig. 2D) also followed the leads of F3 and F4, looped around electrodes Fp1 and Fp2, respectively, then around the Gnd, and connected to the Ref through the resistors. The resistors were non-magnetic non-inductive surface mount thin film resistors (Vishay PNM1206-50KBCT-ND, 50k, 0.1%, 0.4W). After the cap had been placed and aligned on a participant's head, the wire contours were fixed tightly to the cap's fabric with adhesive tape (3M Durapore). This EEG cap modification enabled acquisition of the four reference artifact waveforms, along with 27 EEG waveforms and one ECG waveform, using the standard 32-channel system configuration. Importantly, the use of the modified EEG cap did not affect quality of structural or functional MRI brain images.

## S1.2. Performance of real-time EEG artifact regression

Figure S1 evaluates effectiveness of the real-time regression of the CB and motion artifacts (Sec. 2.5, Fig. 3) with the linear regression coefficients $\{a_i\}$, $\{b_i\}$, $i=1...4$, based on the previous run as opposed to the current run. The actual real-time signal variance changes for channels F3 and F4 (same as in Fig. 3C) after the regression procedure with the coefficients for a current run determined by fitting the data for the previous run are shown along the y-axis in Fig. S1. The corresponding variance changes after an offline regression procedure with the optimum regression coefficients for a current run are shown along the x-axis. The results are pooled across five task runs for all participants. The plots in Fig. S1 demonstrate high correlation ($r=0.98$) between the two quantities. A few outlier points (included in the statistics computation) out of total $n=120$ points in each plot correspond to situations when a current run is characterized by a very drastic (usually one per run) head motion, not present during the previous run, or vice versa. The linear fits to the data for F3 and F4 have slopes of 0.96 and 0.98, respectively (Fig. S1). Therefore, the use of the regression coefficients optimized for the previous run yields signal variance reduction that is 96-98%, on the average, of the variance reduction achieved with the regression coefficients optimized for the current run. These results justify our proposed implementation of the linear regression procedure for improved real-time EEG-fMRI artifact correction, described in Sec. 2.5.

## S1.3. Reliability of the real-time EEG artifact correction

Figure S2 compares the EEG-nf target measures computed in real time and the corresponding measures determined in offline EEG data analysis. The relative frontal EEG asymmetries $A$ and $B$ were computed every 2 s (Sec. 2.6) following the real-time EEG-fMRI artifact correction procedure (Sec. 2.5), and their values were saved to a file during each run. The real-time $A$ and $B$ data, reported in Fig. S2, were taken from these files. The offline EEG data processing was applied to the raw EEG data recorded during fMRI. It included average artifact subtraction (AAS) for MR artifacts, followed by AAS for CB artifacts, and exclusion of bad intervals, as described in detail below (S1.6). The offline $A$ and $B$ data in Fig. S2 were computed after such processing.

Mean real-time $A$ values for the Happy Memories conditions in each run and the corresponding mean offline $A$ values are compared in Fig. S2A. The results are pooled across five task runs for all participants. Fig. S2B compares mean real-time $A$ changes and mean offline $A$ changes between the Rest and Happy Memories conditions. Similar plots for the $B$ values and $B$ changes are shown in Figs. S2C and S2D. The correlations in Figs. S2A-D are highly significant with large effect sizes ($r>0.5$). These results suggest that the real-time EEG-nf target measures were sufficiently reliable, both for the alpha band ($A$) and for the high-beta band ($B$). The differences between the real-time and offline data in Fig. S2 can be attributed to the following factors. First, the offline AAS procedures for MR and CB artifacts are more accurate than the corresponding real-time AAS procedures, because they involve careful visual inspection of the artifact patterns across an entire run and manual adjustment of the correction parameters (semi-automatic correction mode). Second, the real-time artifact regression procedure (Fig. 3B) can reduce some CB artifacts that cannot be efficiently suppressed by the offline AAS, e.g. due to large variations in CB artifacts' temporal profiles.

*S1.4. fMRI data analysis*

Offline analysis of the fMRI data was performed in AFNI (Cox, 1996; Cox and Hyde, 1997). Pre-processing of single-subject fMRI data included time series despiking using the 3dDespike AFNI program with -localedit option. It was followed by correction of cardiorespiratory artifacts using the AFNI implementation of the RETROICOR method (Glover et al., 2000). Further fMRI pre-processing involved slice timing correction and volume registration of all EPI volumes acquired in the experiment using the 3dvolreg AFNI program with two-pass registration. The last volume of the short EPI dataset, acquired immediately after the high-resolution anatomical MPRAGE brain image (Sec. 2.6), was used as the registration base.

To enable transformation of the fMRI data to the Talairach space, the Talairach transform was first performed for each subject's high-resolution anatomical MPRAGE brain image. The image was subjected to explicit skull-stripping using the 3dSkullStrip AFNI program with -blur_fwhm option, and then transformed towards the standard TT_N27 template in the Talairach space using the @auto_tlrc AFNI program.

The fMRI activation analysis was performed according to the standard general linear model (GLM) approach. It was conducted for each of the five task fMRI runs (Fig. 1B) using the 3dDeconvolve AFNI program. The GLM model included two block-design stimulus condition terms, Happy Memories and Count, represented by the standard block-stimulus regressors in AFNI. A general linear test (-gltsym) term was included to compute the Happy vs Count contrast. Nuisance covariates included the six fMRI motion parameters and five polynomial terms for modeling the baseline. To further reduce effects of residual motion artifacts, the fMRI data and motion parameters were lowpass Fourier filtered at 0.1 Hz prior to the GLM analysis. GLM $\beta$ coefficients were computed for each voxel, and average percent signal changes for Happy vs Rest, Count vs Rest, and Happy vs Count contrasts were obtained by dividing the corresponding $\beta$ values ($\times 100\%$) by the $\beta$ value for the constant baseline term. The resulting fMRI percent signal change maps for each run were transformed to the Talairach space by means of the @auto_tlrc AFNI program. The individual high-resolution anatomical brain image in the Talairach space was used as the transformation template. The maps were re-sampled to $2\times 2\times 2$ mm$^3$ isotropic voxel size.

Average individual BOLD activity levels were computed in the offline analysis for the LA and L rACC target ROIs, exhibited in Figs. 2A,B. The voxel-wise fMRI percent signal change data from the GLM analysis, transformed to the Talairach space, were averaged within these ROIs and used as GLM-based measures of these regions' BOLD activities.

To compare BOLD activity levels for the left and right amygdala, we considered amygdala BOLD laterality, i.e. a difference in mean GLM-based fMRI percent signal changes between the LA and RA ROIs for each contrast, run, and participant. The LA and RA ROIs in this case were defined as the left and right amygdala regions specified in the AFNI implementation of the Talairach-Tournoux brain atlas. Similarly, we computed BOLD laterality for the middle frontal gyrus (MidFG) as a difference in mean GLM-based fMRI percent signal changes between the left and right MidFG ROIs. The ROIs were defined as the left and right middle frontal gyrus regions specified in the AFNI implementation of the Talairach-Tournoux atlas, and limited to a selected slab along $z$-axis ($z_1 \leq z \leq z_2$).

*S1.5. fMRI-based PPI analysis*

To evaluate changes in the left amygdala fMRI functional connectivity between experimental conditions, we conducted a psychophysiological interaction (PPI) analysis (Friston et al., 1997; Gitelman et al., 2003). The analysis was based on fMRI time course for an LA seed ROI. The seed ROI was defined as the left amygdala region specified in the AFNI implementation of the Talairach-Tournoux brain atlas. (We used the anatomical amygdala ROI, because the spherical LA target ROI (Fig. 2A) includes some voxels outside the amygdala proper). The seed ROI was transformed to each subject's individual EPI space. In addition, 10-mm-diameter ROIs were defined within the left and right frontal white matter (WM) and within the left and right ventricle cerebrospinal fluid (CSF) using the individual high-resolution anatomical brain image in the Talairach space (S1.4), and also transformed to the EPI space. The pre-processed fMRI data and the six fMRI motion parameters were bandpass filtered between 0.01 and 0.08 Hz using the 3dTproject AFNI program. The 3dmaskave AFNI program was then used to compute average fMRI time courses for the LA, WM, and CSF ROIs. The LA seed ROI time course was employed as the fMRI-based PPI correlation regressor. A PPI interaction regressor was defined for the Happy Memories vs Rest condition contrast as follows. A [Happy−Rest] contrast function was defined to be equal +1 for the Happy Memories condition blocks, −1 for the preceding Rest condition blocks, and 0 for all other condition blocks (Fig. 1B). The LA time course was detrended, using the 3dTproject AFNI program, with respect to the time courses of the six fMRI motion parameters (together with the same time courses shifted by one *TR*), the time courses for the WM and CSF ROIs, and five polynomial terms. It was then deconvolved using the 3dTfitter AFNI program to estimate a time course of the underlying neuronal activity. This estimated 'neuronal' time course was multiplied by the [Happy−Rest] contrast function, and convolved with the same hemodynamic response function (HRF, 'Cox special') using the waver AFNI program. The resulting waveform was employed as the fMRI-based PPI interaction regressor for the Happy Memories vs Rest condition contrast.

A single-subject fMRI-based PPI analysis involved fitting a GLM model with the two PPI regressors using the 3dDeconvolve AFNI program. The GLM design matrix for each task run included four stimulus regressors, sixteen covariates of no interest, and five polynomial terms for modeling the baseline. The stimulus regressors included the fMRI-based PPI interaction regressor, the fMRI-based PPI correlation regressor, the Happy Memories block-stimulus regressor, and the Count block-stimulus regressor. The last two regressors were the standard block-design fMRI regressors in AFNI corresponding to the stimulus waveforms with 40-s-long condition blocks (Fig. 1B). The covariates of no interest included time courses of the six fMRI motion parameters, time courses of the same parameters shifted by one *TR*, time courses of the left and right WM ROIs, and time courses of the left and right ventricle CSF ROIs. Each single-subject PPI analysis produced GLM-based $R^2$-statistics and *t*-statistics maps for the fMRI-based PPI interaction and correlation terms for each run. These statistics were used to compute voxel-wise PPI interaction and correlation values. The resulting maps were subjected to the Fisher *r*-to-*z* normalization, transformed to the Talairach space, re-sampled to 2×2×2 mm³ isotropic voxel size, and spatially smoothed using isotropic Gaussian blur with FWHM = 5 mm. The single-subject fMRI-based PPI interaction maps were submitted to whole-brain group PPI analyses that employed the 3dttest++ AFNI program.

*S1.6. EEG data analysis*

Offline analysis of the EEG data was performed in BrainVision Analyzer 2.1 (Brain Products, GmbH). Removal of MR and cardioballistic (CB) artifacts was based on the average artifact subtraction (AAS) method (Allen et al., 1998, 2000) implemented in the Analyzer. The MR artifact template was defined using MRI slice markers recorded with the EEG data. After the MR artifact removal, the data were bandpass filtered between 0.5 and 80 Hz (48 dB/octave) and downsampled to 250 S/s sampling rate (4 ms interval). The fMRI slice selection frequency (17 Hz) and its harmonics were removed by band rejection filtering. The MR artifact removal was performed for the 27 EEG channels, the ECG channel, and the 4 reference artifact channels (re-purposed FC1, FC2, TP9, TP10). Removal of CB artifacts and follow-up analyses were conducted for the 27 EEG channels only. The CB artifact template was determined from the cardiac waveform recorded by the ECG channel, and the CB artifact to be subtracted was defined, for each EEG channel, by a moving average over 21 cardiac periods. Cardiac periods with strong random-motion artifacts were not included in the CB correction.

Following the MR and CB artifact removal, the EEG data from the five task runs (Fig. 1B) were concatenated to form a single dataset. The data were carefully examined, and intervals exhibiting significant motion or instrumental artifacts were marked manually as "bad intervals" and excluded from the analysis. The signals from the four reference artifact channels were taken into account to more reliably identify data intervals affected by random head motions and distinguish them from intervals exhibiting neuronal activity (e.g. theta). Channel FCz was kept as the EEG reference throughout the analysis.

An independent component analysis (ICA) was performed over the entire dataset with exclusion of the bad intervals. This approach ensured that independent components (ICs) corresponding to various artifacts were identified and removed in a consistent manner across all five runs. The Infomax ICA algorithm (Bell and Sejnowski, 1995), implemented in BrainVision Analyzer 2.1, was applied to the data from 27 EEG channels and yielded 27 ICs. Time courses, spectra, topographies, and kurtosis values of all the ICs were carefully analyzed to identify various artifacts, as well as EEG signals of neuronal origin. After all the ICs had been classified, an inverse ICA transform was applied to remove the identified artifacts from the EEG data. Because many artifacts had already been removed using the ICA, the data were examined again, and new bad intervals were defined to exclude remaining artifacts.

A time-frequency analysis was performed to compute EEG power for each channel as a function of time and frequency. The continuous wavelet transform with Morlet wavelets, implemented in BrainVision Analyzer 2.1, was applied to obtain EEG signal power in [0.5-30] Hz frequency range with 0.5 Hz frequency resolution and 4 ms temporal sampling. An EEG power as a function of time was then computed for each band of interest.

*S1.7. EEG-based PPI analyses*

To investigate how temporal correlations between FAA (or FBA) and BOLD activity changed between experimental conditions, we performed PPI analyses adapted for EEG-fMRI (Zotev et al., 2014, 2016, 2018a). The analyses followed the standard fMRI-based PPI analysis approach (Friston et al., 1997; Gitelman et al., 2003), except that the initial deconvolution step, used to estimate an underlying neuronal activity from an fMRI time course, was skipped, and the actual EEG activity time course was employed. The EEG-based PPI analyses were conducted separately for the FAA and FBA time courses.

EEG-based PPI regressors for the FAA were defined as illustrated in Figure S3. The FAA values, computed with 4 ms temporal resolution for each experimental run, were averaged for 200-ms-long time bins. The resulting waveform was linearly detrended and orthogonalized with respect to the Happy Memories and Count stimulus waveforms (Fig. 1B) using the glmfit() MATLAB program. This procedure removed variations in mean FAA levels across the conditions to focus the analysis on temporal FAA variations around the means. The FAA time course was then converted to *z*-scores across each run. The HRF ('Cox special') was calculated with 200 ms sampling using the waver AFNI program. Convolution of the *z*(FAA) time course with the HRF by means of the conv() MATLAB program yielded a regressor, which we employed as the EEG-based PPI correlation regressor (Fig. S3A). An EEG-based PPI interaction regressor was defined for the Happy Memories vs Count condition contrast as follows. A [Happy−Count] contrast function was set to be equal +1 for the Happy Memories condition blocks, −1 for the Count condition blocks, and 0 for the Rest condition blocks (Fig. S3B). The *z*(FAA) time course was first multiplied by the [Happy−Count] contrast function, and then convolved with the HRF (Fig. S3C). The resulting waveform was used as the EEG-based PPI interaction regressor for the Happy Memories vs Count condition contrast. The two PPI regressors – correlation and interaction – were sub-sampled to the middle time points of fMRI volumes. Two additional PPI regressors were defined in the same way using the power-sum function instead of the FAA. Prior to its inclusion in the GLM model, the FAA-based PPI correlation regressor was linearly detrended and orthogonalized with respect to the corresponding power-sum-based PPI correlation regressor. Similarly, the FAA-based PPI interaction regressor was linearly detrended and orthogonalized with respect to the power-sum-based PPI interaction regressor. This procedure ensured that the two EEG-based PPI regressors specifically reflect temporal variations in the FAA rather than variations in the average power for the two channels. For the FBA, the PPI regressors were defined in a similar way starting with the FBA time course.

A single-subject EEG-based PPI analysis involved fitting a GLM model using the 3dDeconvolve AFNI program. The design matrix for each task run had the same structure as described above for the fMRI-based PPI analysis, except that the two EEG-based PPI regressors – interaction and correlation – were used instead of the fMRI-based PPI regressors. The resulting PPI interaction and correlation maps were subjected to the Fisher normalization, Talairach transform, re-sampling, and spatial smoothing as described above. Whole-brain group PPI analyses were conducted for the EEG-based PPI interaction maps using the 3dttest++ AFNI program. The analyses included two covariates: the participants' MADRS depression severity ratings and average individual values of the PPI interaction effect for a WM mask. The WM mask was defined for each participant as follows. The individual high-resolution anatomical brain image in the Talairach space (S1.4) was thresholded to select WM regions only, and the resulting mask was multiplied by the standard WM mask in the Talairach space (TT_wm+tlrc). The mask was then re-sampled to $2\times2\times2$ mm$^3$ voxels, and subjected to erosion by one voxel to improve its separation from gray matter. The resulting individual WM mask in the Talairach space contained, on average, ~10000 voxels. The EEG-based PPI interaction values, averaged within this WM mask, were used as a covariate vector in the group analyses to better account for spurious PPI interaction effects. Such effects could be caused, e.g., by residual motion artifacts in the simultaneously acquired EEG and fMRI data.

*S2.1. Verbal self-report performance ratings*

Figure S4 exhibits average memory recall and happiness ratings for each group. The two ratings were reported verbally by each participant after each experimental run (except Rest). The EG vs CG group differences in the individual ratings averaged across the four nf runs trended toward significance after correction both for the memory recall ratings (EG vs CG, NF: $t(22)=2.29$, $p<0.032$, $q<0.064$, $d=0.99$) and the happiness ratings (EG vs CG, NF: $t(22)=1.90$, $p<0.071$, $q<0.071$, $d=0.82$). For the Transfer run, group differences were significant after correction for both ratings (Memory-recall, EG vs CG, TR: $t(22)=2.61$, $p<0.016$, $q<0.043$, $d=1.13$; Happiness, EG vs CG, TR: $t(22)=2.39$, $p<0.026$, $q<0.043$, $d=1.04$; corrected for the four tests).

*S2.2. BOLD fMRI activity across the brain*

Figure S5 exhibits whole-brain statistical maps of BOLD fMRI activity during the rtfMRI-EEG-nf training for the EG participants. The maps correspond to the Happy Memories vs Count condition contrast (H vs C). The individual-subject fMRI percent signal change maps were averaged for the four nf runs (PR, R1, R2, R3). The group mean was compared to zero using a one-sample *t*-test ($df=15$, two-tailed). The statistical results are summarized in Table S2. The maps in Fig. S5 are FDR corrected with $q<0.05$ threshold, and the data in Table S2 – with $q<0.01$ threshold. The results demonstrate significant positive BOLD activity contrast for the left amygdala region and many areas of the limbic system. They also reveal pronounced BOLD laterality for large parts of the middle frontal gyrus and superior frontal gyrus (Fig. S5, Table S2).

## S2.3. Amygdala fMRI connectivity changes

Figure S6 shows whole-brain statistical maps for the EG vs CG group difference in the LA fMRI connectivity changes during the rtfMRI-EEG-nf training. The statistics are summarized in Table S3. The results reveal three loci in the rACC area, characterized by the most pronounced EG vs CG group differences: (−8, 34, 7) with $t=4.41$, (−9, 41, 5) with $t=4.04$, and (3, 35, 9) with $t=3.34$ (Table S3). These loci are pointed by green arrows in Fig. S6B. The EG vs CG group differences in fMRI connectivity changes between the LA and 10-mm-diameter spherical ROIs centered at these locations were significant with large effect sizes: $t(22)=3.31$, $p<0.003$, $d=1.43$ for the (−8, 34, 7) centered ROI; $t(22)=3.31$, $p<0.003$, $d=1.43$ for the (−9, 41, 5) centered ROI; and $t(22)=3.06$, $p<0.006$, $d=1.32$ for the (3, 35, 9) centered ROI. The average connectivity changes between the LA and the right rACC ROI centered at (3, 35, 9) showed significant negative correlation with the MADRS depression severity ratings for the EG ($r=−0.55$, $p<0.027$).

## S2.4. Amygdala BOLD laterality interpretation

The amygdala BOLD laterality ('LA−RA') is a relevant metric for assessing target-specific effects of the rtfMRI-nf procedure aimed at upregulating the LA BOLD activity for the following reasons (Zotev et al., 2016). First, both the LA and RA are activated by happy emotion induction, as evidenced by substantial BOLD activations of the RA in our studies that used rtfMRI-nf of the LA activity (e.g. Zotev et al., 2011, 2016; Young et al., 2014). Second, emotional side effects that accompany performance of the difficult neurofeedback task (e.g. confusion, frustration, anxiety, etc.) conceivably involve activations of both the LA and RA. Third, due to the relative proximity of the LA and RA, their apparent BOLD activity levels may be affected by similar fMRI artifacts (e.g. signal losses due to magnetic susceptibility variations). Because only the LA BOLD activity (and not that of the RA) is explicitly upregulated using the rtfMRI-nf in our studies, the amygdala BOLD laterality is more sensitive to *target-specific* effects of the rtfMRI-nf procedure.

## S2.5. fMRI correlates of the FBA modulation

Figure S8 shows whole-brain statistical maps for the FBA-based PPI interaction effect for the Happy Memories vs Count contrast for the EG. The group analysis was conducted in the same way as described for the FAA. Statistical results for the FBA-based PPI interaction effect are summarized in Table S4. The maps in Fig. S8 are voxel-wise FDR corrected at $q<0.07$ level, and the data in Table S4 – at $q<0.05$ level.

Similar to the results for the FAA (Fig. 11), the whole-brain maps of the FBA-based PPI interaction effect (Fig. S8) demonstrate that performance of the rtfMRI-EEG-nf task was associated with enhancement in temporal correlations between the FBA and BOLD activities of the large brain network. In particular, significant positive FBA-based PPI interaction effects are observed for the corresponding left and right DLPFC regions (MidFG, BA 9), with maxima at (−29, 35, 32) and (31, 41, 28), respectively (Fig. S8, Table S4). These maxima are located closer to the medial plane, than the similar maxima for the FAA-based PPI effect (Fig. 11, Table 2).

The results in Fig. S8 show the most pronounced FBA-based PPI interaction effects along the cortical midline and the cingulate gyrus, including the anterior and posterior cingulate (Fig. S8, Table S4). The statistical maximum is at (−19, 12, 42) in the left SFG (BA 8) and the effect extends down to the ACC (BA 32). Elevated high-beta activity in these areas, often with some lateralization to the right, is associated with anxiety (e.g. Zotev and Bodurka, 2020 and references therein). Significant FBA-based PPI interaction effects are also found for many regions involved in autobiographical memory retrieval, including the hippocampus, the extended areas of the parahippocampal gyrus, the anterior thalamus, the precuneus (BA 31), the posterior cingulate (BA 29), the lingual gyrus (involved in visual memory), and others (Fig. S8, Table S4). For the PMC areas (PrecG, BA 6), the FBA-based PPI interaction effects are less pronounced than the corresponding FAA-based effects.

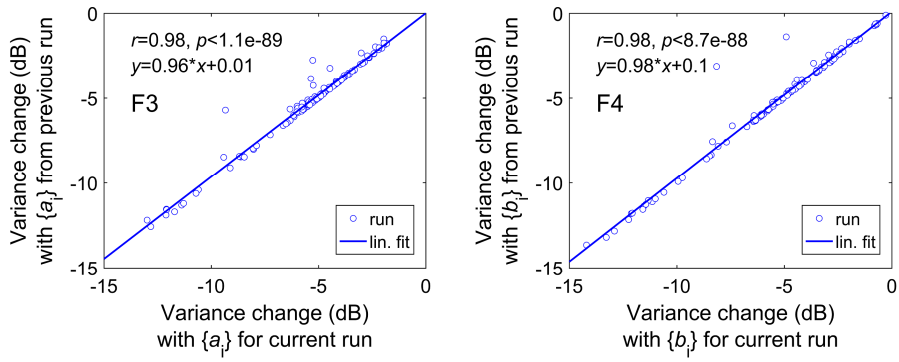

**Figure S1.** Effectiveness of the real-time EEG-fMRI artifact regression procedure with the regression coefficients {$a_i$} and {$b_i$} determined by fitting the data for the previous run (y-axis), compared to the effectiveness of a similar procedure with the optimum coefficients determined by fitting the data for the same (current) run (x-axis). Results for 120 experimental runs (24 participants, 5 task runs) are pooled together in each plot.

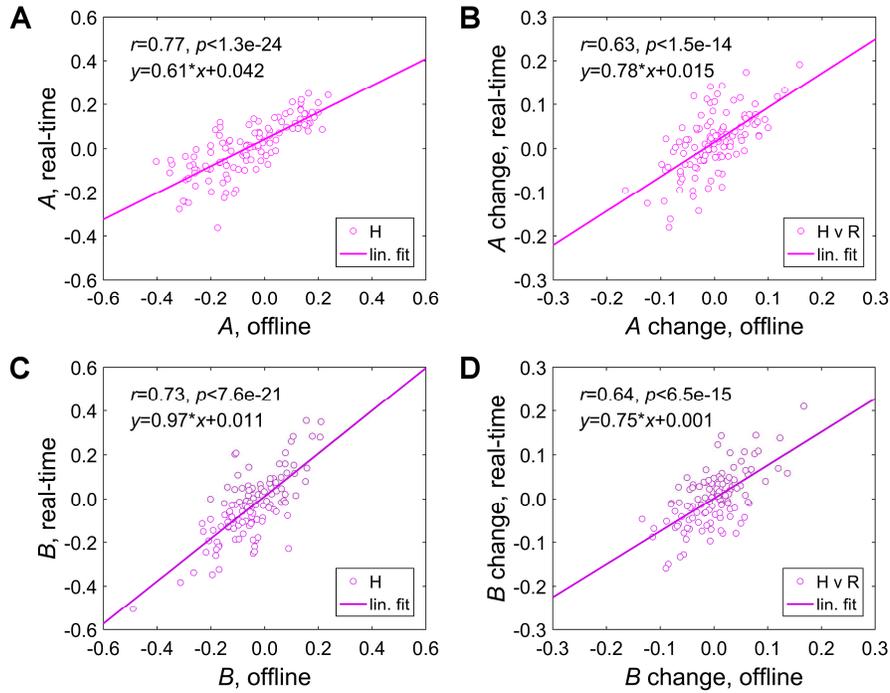

**Figure S2.** Comparison of relative frontal EEG asymmetry measures *A* (alpha band) and *B* (high-beta band), determined for the EEG data after the real-time processing and for the same data after the offline EEG data processing. Results for 120 experimental runs (24 participants, 5 task runs) are pooled together in each plot. A) Mean *A* values for the Happy Memories conditions in each run. B) Mean *A* changes between the Rest and Happy Memories conditions (H vs R) in each run. C) Mean *B* values for the Happy Memories conditions. D) Mean *B* changes between the Rest and Happy Memories conditions.

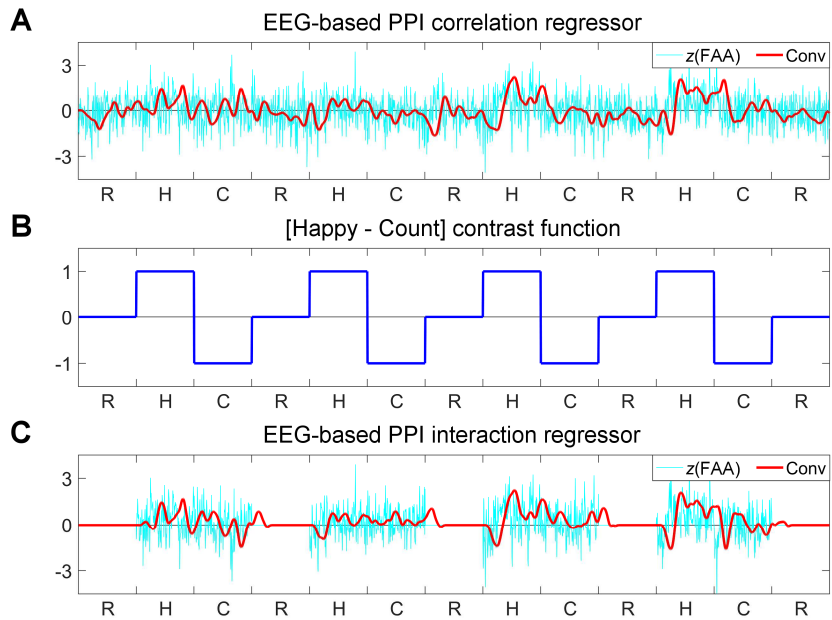

**Figure S3.** Definition of regressors for EEG-based psychophysiological interaction (PPI) analysis of fMRI data. A) Convolution of a time course of the FAA (converted to *z*-scores) with the HRF yields an EEG-based PPI correlation regressor. B) Contrast function for the Happy Memories (H) versus Count (C) conditions for one experimental run. C) Convolution of the FAA time course, multiplied by the contrast function, with the HRF yields an EEG-based PPI interaction regressor for the Happy Memories vs Count condition contrast.

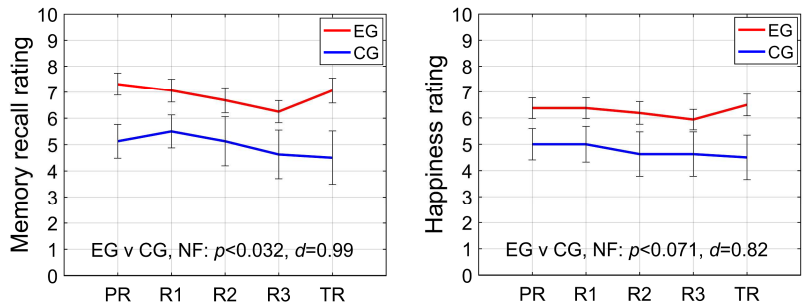

**Figure S4.** Average memory recall and happiness ratings reported by the participants after each experimental run. EG – experimental group, CG – control group. The error bars are standard errors of the mean (sem). The NF refers to group difference statistics (*p*-value from an independent-samples *t*-test and the corresponding effect size *d*) for the individual ratings averaged across the four nf runs (PR, R1, R2, R3).

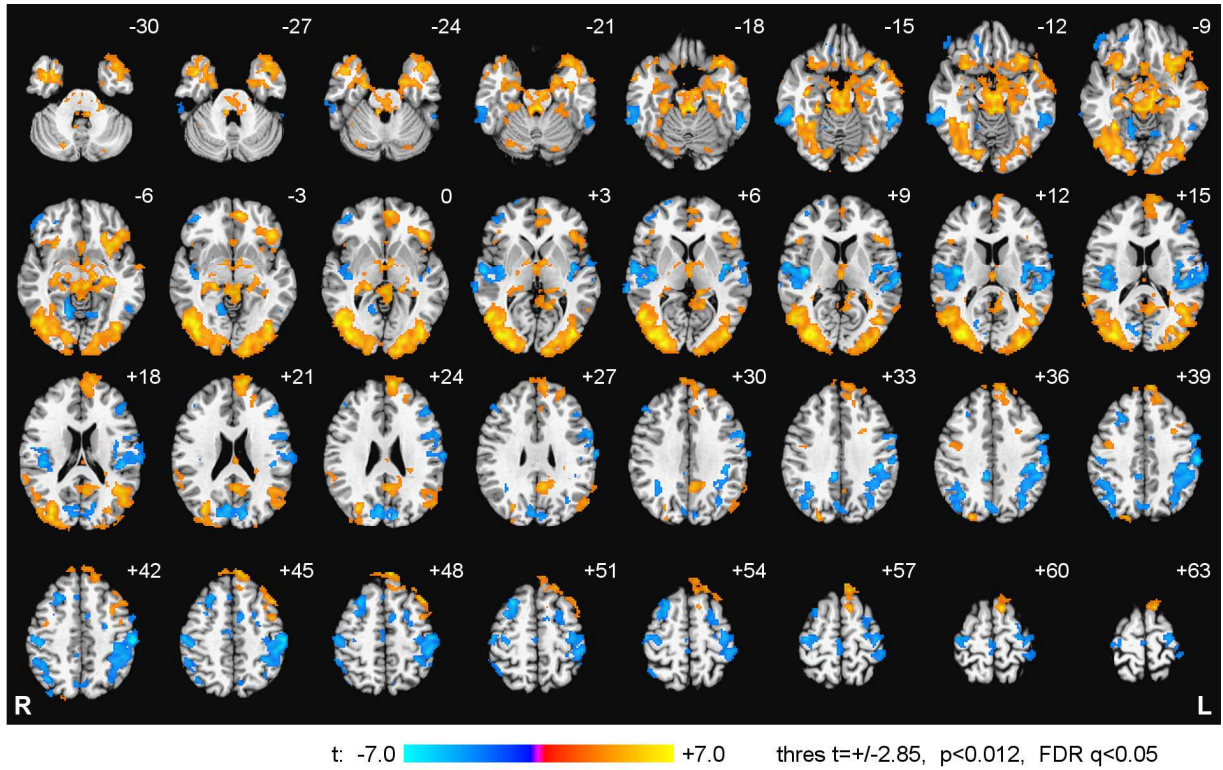

**Figure S5.** Statistical maps of BOLD fMRI activity, corresponding to the Happy Memories vs Count condition contrast, during the rtfMRI-EEG-nf training for the experimental group (EG). The individual fMRI activity results were averaged across the four nf runs (PR, R1, R2, R3). The maps are voxel-wise FDR corrected and projected onto the standard TT_N27 anatomical template in the Talairach space, with 3 mm separation between axial slices. The number adjacent to each slice indicates the *z* coordinate in mm. Following the radiological notation, the left hemisphere (L) is shown to the reader's right. The *t*-statistics maxima and minima and the corresponding locations are specified in Table S2.

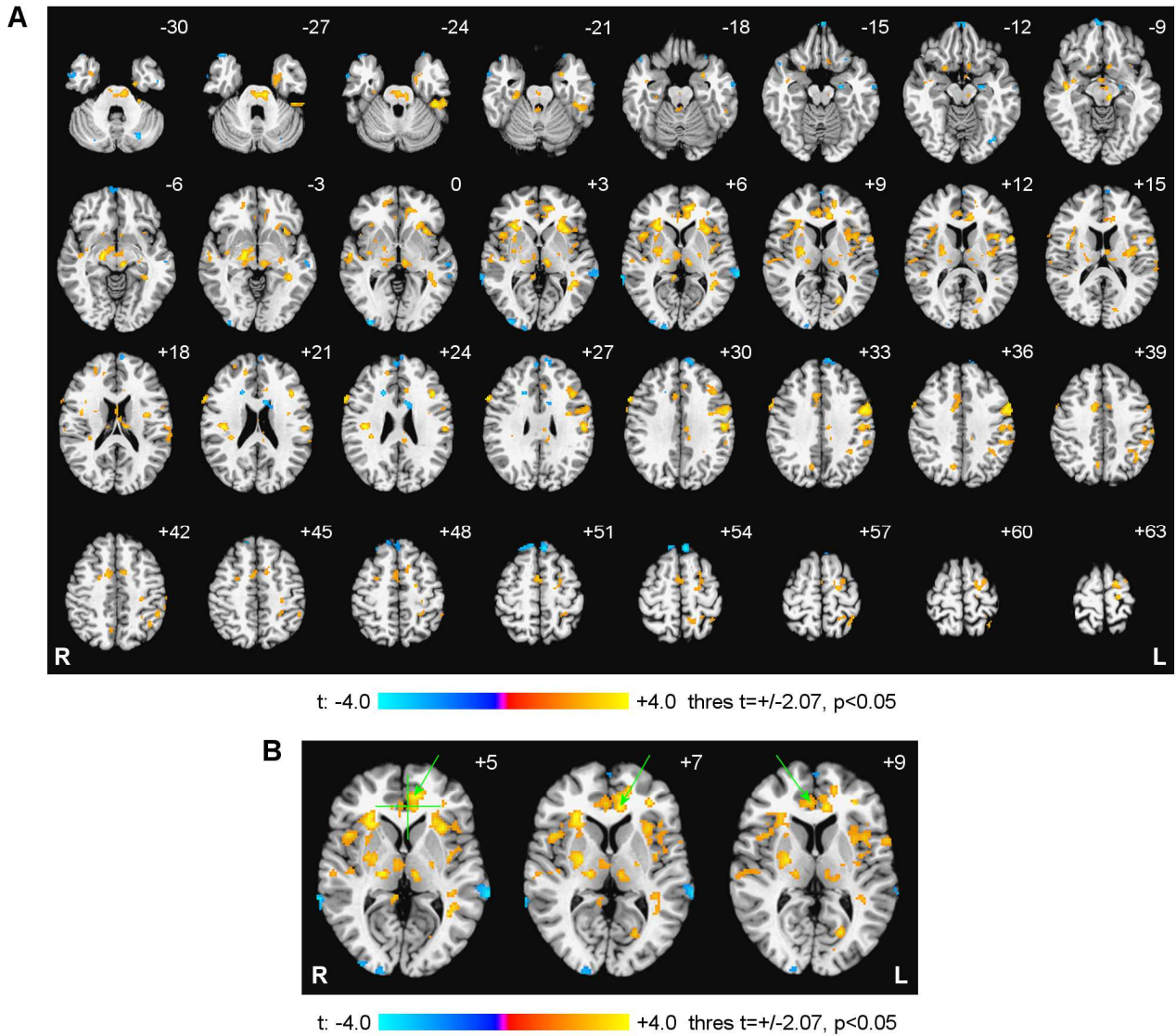

**Figure S6.** A) Statistical maps of the experimental vs control group difference (EG vs CG) in the left amygdala fMRI functional connectivity changes during the rtfMRI-EEG-nf training. The fMRI connectivity changes between the Rest and Happy Memories conditions (H vs R) were evaluated in the psychophysiological interaction (PPI) analysis, based on the LA time course. The individual PPI interaction results were averaged across the four nf runs (PR, R1, R2, R3). The maps are projected onto the standard TT_N27 anatomical template in the Talairach space. The number adjacent to each slice indicates the *z* coordinate in mm. Peak *t*-statistics values and the corresponding locations are specified in Table S3. B) Loci in the rACC area, which exhibited the largest EG vs CG group differences, are pointed by green arrows: (−8, 34, 7), (−9, 41, 5), and (3, 35, 9). The green crosshairs mark the center of the L rACC target ROI.

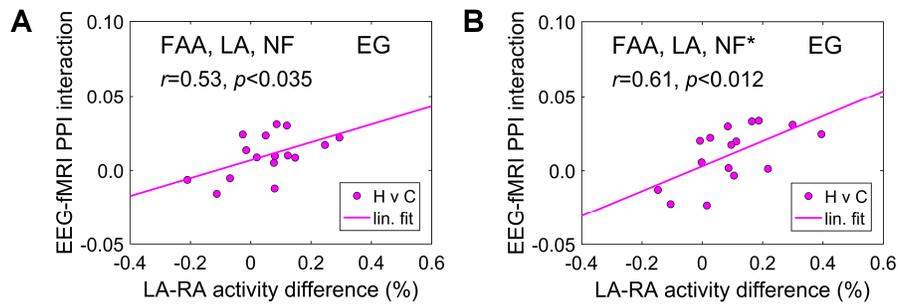

**Figure S7.** Correlations between average values of the amygdala BOLD laterality ('LA−RA') and the corresponding average values of the FAA-based PPI interaction effect for the LA ROI for the experimental group (EG). Both quantities correspond to the Happy Memories vs Count (H vs C) condition contrast. A) The individual results were averaged across the four nf runs (PR, R1, R2, R3). B) The individual results were averaged for three nf runs (out of four) with the most positive amygdala BOLD laterality values.

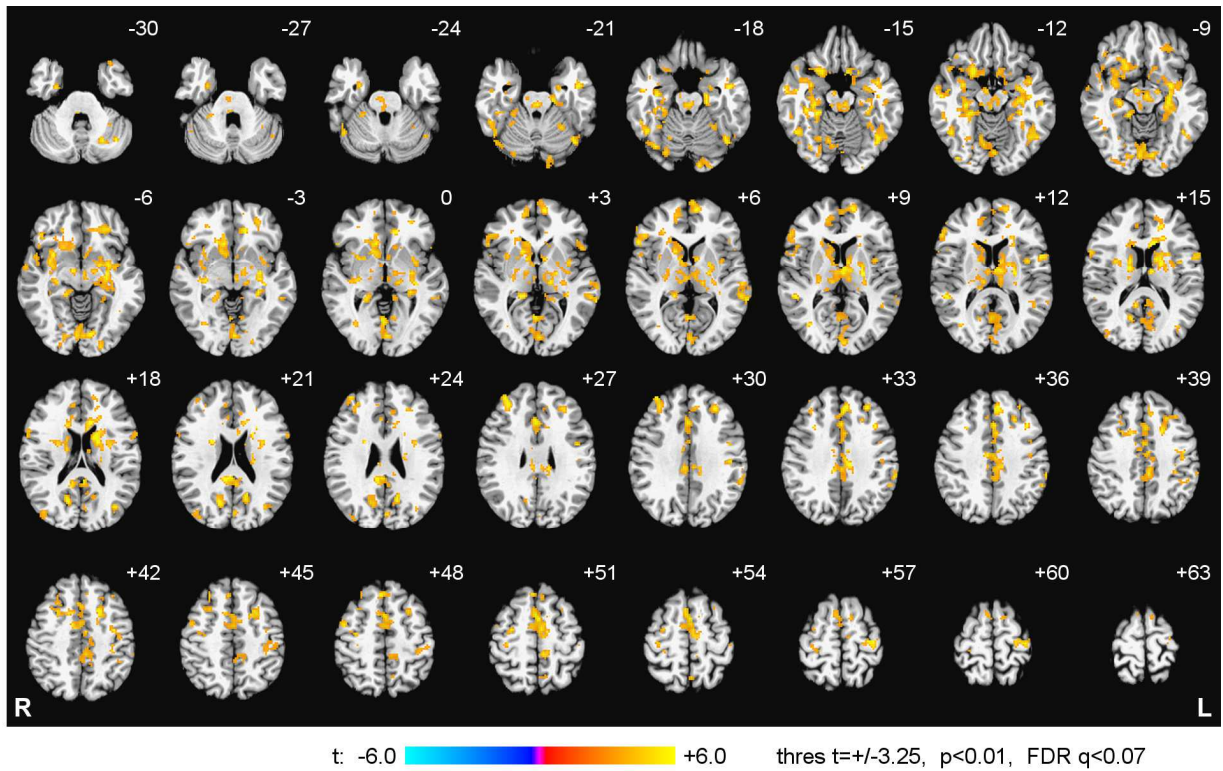

**Figure S8.** Enhancement in temporal correlation between frontal high-beta EEG asymmetry (FBA) and BOLD activity during the rtfMRI-EEG-nf training. Statistical maps of the FBA-based PPI interaction effect for the Happy Memories vs Count condition contrast (H vs C) are shown for the experimental group (EG). The maps are voxel-wise FDR corrected and projected onto the standard TT_N27 anatomical template, with 3 mm separation between axial slices. The number adjacent to each slice indicates the $z$ coordinate in mm. The left hemisphere (L) is to the reader's right. Peak $t$-statistics values for the FBA-based PPI interaction effect and the corresponding locations are specified in Table S4.

**Table S1.** Psychological trait measures for the study participants. Psychological traits were assessed before the rtfMRI-EEG-nf session using the Hamilton Depression Rating Scale (HDRS), the Montgomery-Asberg Depression Rating Scale (MADRS), the Snaith-Hamilton Pleasure Scale (SHAPS), the Hamilton Anxiety Rating Scale (HARS), the State-Trait Anxiety Inventory (STAI), the Toronto Alexithymia Scale (TAS-20), and the Behavioral Inhibition System / Behavioral Activation System scales (BIS/BAS).

| Measure | Experimental group, mean (SD) | Control group, mean (SD) | Difference $t$-score [$p$]# |
|---|---|---|---|
| Participants | 16 | 8 | |
| Age (years) | 32 (11) | 34 (7) | −0.67 [0.510] |
| HDRS | 14.4 (7.0) | 15.1 (4.9) | −0.25 [0.807] |
| MADRS | 19.6 (10.7) | 20.5 (5.7) | −0.23 [0.821] |
| SHAPS | 27.1 (6.7) | 31.5 (5.8) | −1.58 [0.129] |
| HARS | 13.2 (7.5) | 16.1 (6.4) | −0.95 [0.355] |
| STAI Trait anxiety | 56.9 (9.9) | 59.6 (9.6) | −0.65 [0.522] |
| TAS-20 Total alexithymia | 53.5 (14.4) | 61.8 (12.2) | −1.38 [0.180] |
| BAS Reward responsiveness | 14.9 (3.3) | 16.3 (2.4) | −0.99 [0.337] |
| BIS | 22.4 (3.6) | 24.1 (4.1) | −1.02 [0.319] |

# $t(22)$, but $t(19)$ for BIS/BAS; $p$ – two-tailed, uncorrected

**Table S2.** BOLD fMRI activity, corresponding to the Happy Memories vs Count condition contrast, during the rtfMRI-EEG-nf training for the experimental group (EG).

| Region | Laterality | x, y, z (mm) | t-score |
|---|---|---|---|
| **Frontal lobe** | | | |
| Inferior frontal gyrus (BA 47) | L | −33, 17, −18 | 8.26 |
| Middle frontal gyrus (BA 8) | L | −40, 20, 47 | 7.21 |
| Inferior frontal gyrus (BA 47) | L | −43, 29, −5 | 7.09 |
| Superior frontal gyrus (BA 8) | L | −7, 49, 48 | 6.90 |
| Superior frontal gyrus (BA 10) | L | −7, 60, 24 | 6.71 |
| Inferior frontal gyrus (BA 13/47) | R | 29, 14, −10 | 6.69 |
| Superior frontal gyrus (BA 6) | L | −11, 13, 62 | 6.66 |
| Superior frontal gyrus (BA 8) | R | 27, 17, 49 | −6.59 |
| Superior frontal gyrus (BA 6) | L | −9, 31, 57 | 6.36 |
| **Temporal lobe** | | | |
| Middle temporal gyrus (BA 37) | R | 49, −59, 5 | 9.14 |
| Superior temporal gyrus (BA 38) | R | 39, 5, −28 | 7.90 |
| Middle temporal gyrus (BA 37) | L | −42, −67, 11 | 6.65 |
| Inferior temporal gyrus (BA 37) | L | −55, −45, −16 | −5.66 |
| Middle temporal gyrus (BA 20) | R | 53, −35, −15 | −5.35 |
| **Parietal lobe** | | | |
| Postcentral gyrus (BA 2) | L | −55, −23, 44 | −10.9 |
| Precuneus (BA 7/19) | R | 23, −81, 35 | 6.07 |
| Postcentral gyrus (BA 3) | R | 45, −19, 54 | −5.50 |
| Inferior parietal lobule (BA 40) | L | −46, −33, 45 | −5.29 |
| Inferior parietal lobule (BA 39) | R | 35, −63, 42 | −5.24 |
| **Occipital lobe** | | | |
| Middle occipital gyrus (BA 19/18) | R | 39, −81, 2 | 9.70 |
| Middle occipital gyrus (BA 19) | L | −39, −77, 2 | 8.26 |
| Lingual gyrus (BA 17) | L | −19, −86, 1 | 6.69 |
| Cuneus (BA 18) | L | −11, −79, 21 | −5.64 |
| **Limbic lobe** | | | |
| Parahippocampal gyrus (BA 36) | R | 25, −31, −14 | 7.07 |
| Anterior cingulate (BA 32) | L | −7, 49, −2 | 6.62 |
| Amygdala / parahipp. gyrus | L | −29, −1, −14 | 6.55 |
| Amygdala / uncus | L | −21, −3, −22 | 6.13 |
| Cingulate gyrus (BA 31) | R | 7, −39, 34 | −6.10 |
| Hippocampus / parahipp. gyrus | L | −28, −24, −9 | 6.00 |
| Cingulate gyrus (BA 31) | L | −6, −49, 28 | 5.36 |
| **Sub-lobar** | | | |
| Red nucleus | L | −7, −19, −10 | 8.43 |
| Insula (BA 13) | R | 37, −13, 12 | −7.12 |
| Insula (BA 13) | L | −39, −9, 6 | −6.07 |
| Thalamus, mediodorsal | R | 2, −9, 2 | 5.76 |
| Thalamus, mediodorsal | L | −1, −14, 8 | 5.53 |
| Culmen | R | 13, −57, −6 | −5.41 |

FDR $q<0.01$ for $|t|>5.0$; BA – Brodmann areas; L – left; R – right; $x, y, z$ – Talairach coordinates.

**Table S3.** Experimental vs control group difference in the left amygdala fMRI functional connectivity changes between the Rest and Happy Memories conditions (H vs R) during the rtfMRI-EEG-nf training.

| Region | Laterality | x, y, z (mm) | t-score |
|---|---|---|---|
| **Frontal lobe** | | | |
| Inferior frontal gyrus (BA 9) | R | 57, 13, 32 | 5.73 |
| Middle frontal gyrus (BA 9) | L | −53, 8, 36 | 4.63 |
| Superior frontal gyrus (BA 6) | L | −7, 33, 54 | −4.56 |
| Precentral gyrus (BA 6) | L | −51, 1, 33 | 4.49 |
| Middle frontal gyrus (BA 9/46) | L | −39, 19, 24 | 4.42 |
| Superior frontal gyrus (BA 6) | L | −15, −7, 64 | 3.68 |
| Inferior frontal gyrus (BA 44) | L | −55, 5, 14 | 3.67 |
| Medial frontal gyrus (BA 11) | L | −3, 61, −14 | −3.64 |
| Superior frontal gyrus (BA 6) | R | 9, 37, 52 | −3.47 |
| **Temporal lobe** | | | |
| Fusiform gyrus (BA 20) | L | −43, −31, −24 | 5.22 |
| Superior temporal gyrus (BA 22) | L | −67, −35, 6 | −3.82 |
| Superior temporal gyrus (BA 21) | R | 55, −13, 0 | 3.31 |
| **Parietal lobe** | | | |
| Postcentral gyrus (BA 2) | L | −51, −19, 28 | 4.37 |
| Inferior parietal lobule (BA 40) | L | −47, −37, 38 | 3.28 |
| **Occipital lobe** | | | |
| Middle occipital gyrus (BA 18) | R | 31, −91, 2 | −4.01 |
| **Limbic lobe** | | | |
| Anterior cingulate (BA 24) | L | −8, 34, 7 | 4.41 |
| Parahippocampal gyrus (BA 37) | L | −33, −39, −4 | 4.36 |
| Anterior cingulate (BA 32) | L | −9, 41, 5 | 4.04 |
| Cingulate gyrus (BA 24) | R | 11, 1, 38 | 4.01 |
| Uncus (BA 36) | L | −21, −9, −30 | 3.62 |
| Anterior cingulate (BA 24) | R | 3, 35, 9 | 3.34 |
| **Sub-lobar** | | | |
| Subthalamic nucleus | R | 15, −13, −4 | 7.39 |
| Red nucleus | L | −7, −25, −6 | 4.44 |
| Insula (BA 13) | R | 29, 21, 4 | 4.29 |
| Insula (BA 13/47) | L | −31, 15, 4 | 4.15 |
| Thalamus, ventroposterolateral | R | 17, −19, 6 | 3.95 |
| Claustrum | L | −31, −11, 14 | 3.91 |
| Putamen | R | 23, −9, 8 | 3.79 |

$p<0.05$, uncorr. BA – Brodmann areas; L – left; R – right; x, y, z – Talairach coordinates

**Table S4.** Psychophysiological interaction effect, based on the time course of frontal high-beta EEG asymmetry (FBA), for the Happy vs Count contrast for the experimental group (EG).

| Region | Laterality | x, y, z (mm) | t-score |
|---|---|---|---|
| **Frontal lobe** | | | |
| Superior frontal gyrus (BA 8/32) | L | −19, 12, 42 | 11.8 |
| Medial frontal gyrus (BA 6) | L | −5, 35, 36 | 10.2 |
| Superior frontal gyrus (BA 8) | R | 3, 29, 50 | 9.58 |
| Inferior frontal gyrus (BA 44) | L | −51, 6, 13 | 8.75 |
| Medial frontal gyrus (BA 6) | L | −1, 3, 52 | 7.93 |
| Middle frontal gyrus (BA 9) | L | −29, 35, 32 | 7.79 |
| Precentral gyrus (BA 4) | L | −29, −21, 58 | 7.52 |
| Middle frontal gyrus (BA 9) | R | 31, 41, 28 | 7.30 |
| Precentral gyrus (BA 6) | R | 27, −9, 52 | 6.23 |
| Medial frontal gyrus (BA 10) | L | −9, 55, 6 | 6.12 |
| **Temporal lobe** | | | |
| Middle temporal gyrus (BA 39) | L | −37, −69, 18 | 7.91 |
| Superior temporal gyrus (BA 41) | R | 49, −37, 10 | 7.58 |
| Middle temporal gyrus (BA 20) | L | −45, −5, −20 | 7.42 |
| Superior temporal gyrus (BA 22) | L | −63, −41, 6 | 6.88 |
| Middle temporal gyrus (BA 21) | R | 59, −7, −12 | 6.37 |
| **Parietal lobe** | | | |
| Precuneus (BA 31) | L | −13, −59, 24 | 9.79 |
| Precuneus (BA 31) | R | 15, −63, 20 | 8.16 |
| **Occipital lobe** | | | |
| Lingual gyrus (BA 18) | L | −9, −79, −6 | 7.20 |
| **Limbic lobe** | | | |
| Anterior cingulate (BA 33) | L | −9, 19, 16 | 10.3 |
| Posterior cingulate (BA 29) | R | 3, −39, 20 | 9.53 |
| Anterior cingulate (BA 32) | L | −13, 33, −2 | 9.36 |
| Parahippocampal gyrus (BA 30) | R | 17, −37, 4 | 8.40 |
| Parahippocampal gyrus (BA 36) | R | 24, −33, −14 | 8.19 |
| Cingulate gyrus (BA 32) | L | −1, 21, 28 | 7.94 |
| Subcallosal gyrus (BA 47) | R | 20, 11, −12 | 7.65 |
| Cingulate gyrus (BA 31) | L | −10, −25, 34 | 7.61 |
| Hippocampus / parahipp. gyrus | L | −27, −23, −10 | 7.41 |
| Hippocampus / parahipp. gyrus | L | −27, −33, −3 | 7.29 |
| Amygdala / parahipp. gyrus | L | −28, −5, −10 | 6.94 |
| Parahippocampal gyrus (BA 28) | L | −19, −23, −18 | 6.58 |
| **Sub-lobar** | | | |
| Thalamus, ventral anterior | R | 15, −9, 15 | 9.12 |
| Declive | R | 33, −69, −17 | 9.12 |
| Putamen | L | −23, −13, 8 | 9.10 |
| Culmen of vermis | R | 1, −61, 2 | 8.64 |
| Claustrum | R | 29, 7, −6 | 8.14 |
| Caudate body | L | −16, 5, 18 | 7.87 |
| Declive | L | −39, −61, −20 | 7.52 |
| Thalamus, anterior | L | −7, −11, 15 | 7.50 |
| Thalamus, mediodorsal | L | −3, −10, 10 | 7.21 |
| Insula (BA 13) | L | −39, −1, 14 | 6.86 |

FDR $q<0.05$ for $|t|>6.0$; BA – Brodmann areas; L – left; R – right; x, y, z – Talairach coordinates.